\documentclass{jfm}
\usepackage[T1]{fontenc}
\usepackage{lmodern}
\usepackage{graphicx}
\usepackage{subcaption}
\usepackage{booktabs}
\usepackage{multirow}
\usepackage{placeins}
\makeatletter
\def\fps@figure{htbp}
\def\fps@table{htbp}
\makeatother
\usepackage{tikz}
\usetikzlibrary{arrows.meta,decorations.pathmorphing,patterns,calc}
\makeatletter
\def\absfooterflag{}%
\def\pagelimitfooter{}%
\makeatother
\makeatletter
\def\@underjournal{%
  \vbox to 5.5\p@{\noindent
    \parbox[t]{4.5in}{\normalfont\indexsize{\itshape This draft was prepared using the
      LaTeX style file belonging to the Journal of Fluid Mechanics}\\[2.5\p@]
      {\ \ }}%
  \vss}%
}
\def\ps@titlepage{\leftskip\z@\let\@mkboth\@gobbletwo\vfuzz=5\p@
	\def\@oddhead{\vbox{\vspace*{-1pc}%
			\hbox to \textwidth{\@j@urnal \hfil\llap{\thepage}}\par\vskip4pt}}%
	\def\@evenhead{\@j@urnal \hfil\llap{\thepage}}%
	\def\@oddfoot{}%
	\def\@evenfoot{}%
	\def\sectionmark##1{}%
	\def\subsectionmark##1{}%
}
\makeatother
\usepackage{titlesec}
\titleformat{\paragraph}[runin]{\normalfont\bfseries}{\theparagraph}{1em}{}

\titleformat{\paragraph}
{\normalfont\normalsize\bfseries}
{\theparagraph}
{1em}
{}

\titlespacing*{\paragraph}
{0pt}{3.25ex plus 1ex minus .2ex}{1.5ex plus .2ex}

\usepackage{newtxtext}
\usepackage{newtxmath}
\usepackage{natbib}
\usepackage{url}
\usepackage{xr-hyper}
\usepackage{hyperref}
\hypersetup{
	colorlinks = true,
	urlcolor   = blue,
	citecolor  = black,
}
\numberwithin{equation}{section}

\title{Load-induced Taylor-Aris dispersion in compliant electroosmotic pumps with viscoelastic electrolytes}

\author{Subhajyoti Sahoo\aff{1}, 
	Ameeya Kumar Nayak\aff{1}\corresp{\email{ameeyakumar@gmail.com}} 
	\and 
	Kirti Chandra Sahu\aff{2}\corresp{\email{ksahu@che.iith.ac.in}}}

\affiliation{
	\aff{1}Department of Mathematics, Indian Institute of Technology Roorkee, India \\
	\aff{2}Department of Chemical Engineering, Indian Institute of Technology Hyderabad, India
}

\titleformat{\paragraph}[runin]{\normalfont\bfseries}{\theparagraph}{1em}{}
\begin{document}
	\raggedbottom 
	\maketitle
	
	\begin{abstract}
We investigate the dispersion of a passive, neutral solute in a compliant slit, in which electroosmotic flow drives a viscoelastic electrolyte against an outlet pressure load. The electrolyte is modelled using the simplified Phan–Thien–Tanner model with an inviscid Newtonian solvent. Combining lubrication theory, Debye--H\"uckel linearisation and a local elastic-foundation model, we determine the loaded flow and quantify solute dispersion. A central finding is that hydraulic loading fundamentally alters the thin-electric-double-layer (EDL) limit of shear dispersion. It was observed that shear dispersion vanishes under free-flow conditions, but an adverse pressure gradient continues to support core shear under load, resulting in a finite dispersion plateau even without flow reversal. This plateau increases with hydraulic loading and is further enhanced by viscoelasticity at a fixed throughput fraction. For Newtonian flow, wall expansion leaves the plateau unchanged under constant-current conditions but increases it under constant-voltage conditions. At finite EDL thicknesses, opposing electroosmotic and pressure-driven shear can partially cancel, improving the plate number despite reduced throughput. The combined effects of shear dispersion, molecular diffusion and transit time produce an optimal EDL thickness at prescribed throughput. Joint optimisation shifts the optimum towards free-flow conditions and thinner EDLs for the range of parameters considered. The predicted plate numbers agree closely with Brownian-dynamics simulations for six benchmark cases. These results establish the role of hydraulic loading and wall compliance in controlling the trade-off between pressure delivery and band broadening in compliant electroosmotic pumps.

	\end{abstract}
	
	\section{Introduction}\label{sec:intro}
	
	Electroosmotic flow (EOF) enables low-dispersion sample transport due to its nearly uniform core velocity in pressure-free flow through a uniformly charged slit with a thin electric double layer (EDL) \citep{Ghosal2006Electrokinetic,Chatterjee2022Effect}. The axial electric body force per unit volume is the product of the net ionic charge density and the local axial electric field \citep{Masliyah2006electrokinetic}. In EOF against a hydraulic load, the adverse pressure gradient associated with pressure delivery also shears the liquid \citep{Dutta2001Combined}. The physical questions concern the persistence of shear-induced spreading as the EDL thins under load, and the effects of fluid rheology and wall compliance on the balance among pressure delivery, throughput, and solute dispersion.
	
	The hydraulic load couples these transport processes to wall mechanics. Many microfluidic devices are fabricated from compliant materials such as poly(dimethylsiloxane) \citep{McDonald2002Poly,Sollier2011Rapid,Sackmann2014The}. Internal pressure variations can deform such walls, while the resulting changes in channel geometry modify the flow and pressure distribution, producing two-way fluid--structure interaction (FSI) \citep{Chakraborty2012Fluid,Cheung2012In,Kang2014Pressure,Raj2017Hydrodynamics}. Under bulk-Ohmic conduction with uniform conductivity, current conservation couples changes in the channel gap to variations in the axial electric field \citep{Ghosal2002Lubrication}. Previous studies have quantified the dispersion under combined electroosmotic and pressure-driven forcing in rigid channels \citep{Roy2025Dispersion,Samuel2025Taylor}. Here, we examine the effects of pressure gradient, channel gap, and electric field on the thin-EDL dispersion plateau in a loaded compliant slit under constant-current and constant-voltage conditions. For constant-current operation, we also quantify variations in the outlet plate number as a function of the delivered pressure.
	
	In the context of pressure-driven lubrication flows through deformable conduits, \citet{Christov2018Flow} combined lubrication theory with quasi-static plate bending to obtain the nonlinear flow-rate--pressure-drop relation for a shallow rectangular microchannel. Thick-wall elasticity, for which the plate approximation is inadequate, was subsequently considered by \citet{Wang2019Theory}, while \citet{Shidhore2018Static} combined analytical and computational approaches to shallow-channel FSI. Collectively, these studies establish the influence of wall mechanics on the pressure--flow relation in Newtonian pressure-driven flow, providing a basis for evaluating the pressure-dependent channel geometry and its effect on transport. \citet{Rubin2017Elastic} and \citet{Boyko2019Elastohydrodynamics} demonstrated electroosmotic actuation of soft lubricated structures through internal pressure gradients generated by spatially non-uniform EOF. More recently, \citet{Sahoo2026Electroosmotic} considered a Newtonian electrolyte in a constricted channel with a prescribed rigid upper boundary and a pressure-responsive compliant lower wall, including Derjaguin--Landau--Verwey--Overbeek disjoining pressure. These studies demonstrate the coupled effects of electric fields and wall deformation on the pressure response, motivating the evaluation of solute transport using the resulting velocity field.
	
	For viscoelastic electrolytes, polymeric stresses alter the relationship between shear stress and shear rate and can modify the velocity profile, flow rate, and pressure generation \citep{Afonso2009Analytical}, resulting in consequent changes in wall deformation in a compliant channel. For deformable channels, \citet{Boyko2023NonNewtonian} quantified the coupled effects of fluid elasticity and wall compliance on the flow-rate--pressure-drop relation for an Oldroyd--B fluid in the weakly viscoelastic limit. \citet{Mukherjee2022Electrokinetically} examined the effects of wall compliance, spatially varying surface potential and patterned interfacial slip on the pressure distribution and load-bearing capacity of a deformable microchannel. \citet{RamosArzola2021Fluid} treated pressure-driven flow of a simplified Phan-Thien--Tanner (sPTT) fluid using thin- and thick-plate wall models. In a loaded viscoelastic pump, rheology modifies the velocity field and the adverse pressure gradient required to sustain a specified throughput. The effect of rheology on dispersion is therefore evaluated from the transverse variation of the resulting velocity profile.
	
	The seminal work of \citet{Taylor1953Dispersion} and \citet{Aris1956} established the long-time enhancement of longitudinal solute dispersion due to the combined action of shear and transverse molecular diffusion. For rigid slits, \citet{Sadeghi2018Hydrodynamic} analysed hydrodynamic dispersion of viscoelastic EOF, while \citet{Arcos2018Dispersion} quantified the influence of a slowly varying wall zeta potential on dispersion. For shallow flows with heterogeneous electroosmotic mobility, \citet{Dehe2021Hydrodynamic} derived a multiple-scales dispersion model that accounts for the associated internal pressure gradients and benchmarked it against three-dimensional particle tracing. For charged analytes, \citet{Ghosal2012Electromigration} showed that EOF can delay concentration-shock formation and reduce total dispersion over a range of flow strengths, whereas sufficiently strong EOF increases total dispersion through the Taylor--Aris contribution. Charged analytes undergo electrophoretic migration relative to the carrier liquid and can modify the local electric field at concentrations that appreciably alter the electrolyte conductivity. Under electroneutrality and zero-current conditions, \citet{Ding2023Shear} derived an effective long-time transport equation for multispecies electrolytes in channel shear flows, in which unequal ionic diffusivities couple the species fluxes through a diffusion-induced electric field. The neutral solute considered here is transported by advection and molecular diffusion without altering the electrical or hydrodynamic fields.
	
	For passive solutes, \citet{Roy2025Dispersion} analysed neutral-solute dispersion in an sPTT fluid under combined electroosmotic and pressure-driven flow in a rigid microchannel. \citet{Dejam2026Reduced} derived a reduced-order model for viscoelastic mixed-flow dispersion and reported a non-monotonic dependence of the dispersion coefficient on the ratio of pressure-driven to electroosmotic forces. \citet{Wang2026Control} examined dispersion control in polymer solutions under combined magnetic and oscillatory electroosmotic forcing using fractional and classical Maxwell models; that unsteady rigid-tube setting differs from the steady, load-dependent compliant pump considered here. For a rigid Newtonian tube, \citet{Samuel2025Taylor} developed an all-time theory of Taylor dispersion under combined electroosmotic and pressure-driven flow, derived optimal forcing ratios and P\'eclet numbers for minimising dispersion, and benchmarked the transport predictions against Brownian-dynamics simulations. The results of \citeauthor{Samuel2025Taylor} establish that opposing pressure-driven flow can reduce solute dispersion; the corresponding shear balance in a compliant viscoelastic pump also depends on load-induced changes in geometry and electric field. \citet{Jha2026Taylor} derived a macrotransport equation for pressure-driven Newtonian dispersion in a soft axisymmetric channel with a Winkler-type wall response. For prescribed channels with spatially periodic cross-sections, \citet{Ding2025Longtime} derived long-time passive-scalar asymptotics and related the timescale for convergence to the slow manifold to the channel geometry and velocity field. For prescribed axisymmetric channels, \citet{Chang2023Taylor} showed that axial variations in advection can stretch or contract a spatial solute band, so its evolution is not determined by the local dispersion coefficient alone. For channels with slowly varying cross-sectional shape and size, \citet{Samuel2026Varying} derived an area-averaged transport equation and benchmarked solute-dispersion predictions against Brownian-dynamics simulations.
	
	For the compliant electroosmotic pump conveying a viscoelastic electrolyte considered here, the pressure gradient, wall shape and electric field must satisfy the coupled flow, wall-mechanical and current-conservation relations. The sPTT constitutive stress function contains a term linear in the trace of the extra-stress tensor \citep{PhanThien1977New,Afonso2009Analytical}. In the zero-Newtonian-solvent-viscosity limit, the leading-order lubrication relation expresses the shear rate through linear and cubic terms in the total shear stress. The nonzero cubic contribution prevents linear superposition of the local electroosmotic and pressure-driven velocity fields, as established for rigid channels by \citet{Afonso2009Analytical}. Our aim is to characterise the core shear sustained by hydraulic loading as the EDL becomes thin, quantify the effects of rheology and compliance on the residual shear dispersion, and identify conditions under which partial loading increases the plate number relative to free flow despite reducing throughput. The plate number is the square of the mean outlet transit time divided by the variance of that time.
	
	To address these questions, we analyse creeping, isothermal electroosmotic flow and passive-solute transport in a compliant slit microchannel containing an sPTT fluid in the zero-Newtonian-solvent-viscosity limit. Debye--H\"uckel linearisation, lubrication theory, a bulk-Ohmic conduction model with current conservation and a local elastic-foundation law \citep{Skotheim2005Soft,Chandler2020Validity} determine the pressure, wall deformation, electric field and velocity. A closed-form mixed-flow relation retains the pure electroosmotic, pure pressure-driven and nonlinear mixed contributions, and determines pump characteristics under constant-current and constant-voltage operation. We combine a solute-conserving Taylor--Aris reduction with backward first-passage equations to calculate outlet transit-time moments, assuming slow axial variation of the channel geometry and a transverse-diffusion timescale much shorter than the axial transit time. Predictions of mean outlet transit time and plate number under constant-current operation are compared with Brownian-dynamics simulations using the same computed velocity field, thereby assessing the reduced transport description.
	
	At a fixed value of the dimensionless viscoelastic parameter and at a fixed throughput fraction between zero and unity, the dimensionless Taylor--Aris shear-dispersion coefficient approaches a positive limit as the electric double layers thin. The adverse pressure gradient sustains shear in the channel core even when the flow there remains forward. At zero load, the coefficient decreases as the square of the ratio of the Debye length to the undeformed slit half-height. The throughput fraction is the delivered flow rate relative to its zero-load value at the same EDL thickness, rheological parameters, and electrical operating conditions. In the rigid-reference slit, the plateau vanishes quadratically as the throughput fraction approaches unity, with rheology entering via the effective electroosmotic slip. Compliance modifies the loaded velocity profile through coupled changes in the channel gap and electric field, so its influence on the shear-dispersion plateau depends on the electrical operating constraint. At finite EDL thickness, partial cancellation within the combined electroosmotic and pressure-driven shear can reduce the shear-dispersion coefficient below its loaded thin-EDL plateau at the same throughput fraction. In the constant-current cases examined at fixed loaded-throughput fractions, the plate number reaches a maximum at a finite EDL thickness. When both throughput fraction and EDL thickness are varied, the highest plate numbers in the investigated range occur near the free-flow, thin-EDL boundary.
	
	The paper is organised as follows. Section~\ref{sec:formulation} sets out the governing equations for electrohydrodynamic flow, viscoelastic stress, wall deformation and solute transport. Section~\ref{sec:reduction} derives the reduced-order flow and Taylor--Aris transport equations. Section~\ref{sec:num} describes the numerical methods and the convergence checks and compares the solution of the reduced-order transport equations with the Brownian dynamics simulations. Section~\ref{sec:results} presents the pressure--flow characteristics and wall loads and examines solute dispersion and outlet plate number under hydraulic loading. Finally, Section~\ref{sec:concl} summarises the findings and the limits of the model. 
	
	\section{Problem formulation}\label{sec:formulation}
	
	\begin{figure}
		\centering
		\includegraphics[width=\textwidth]{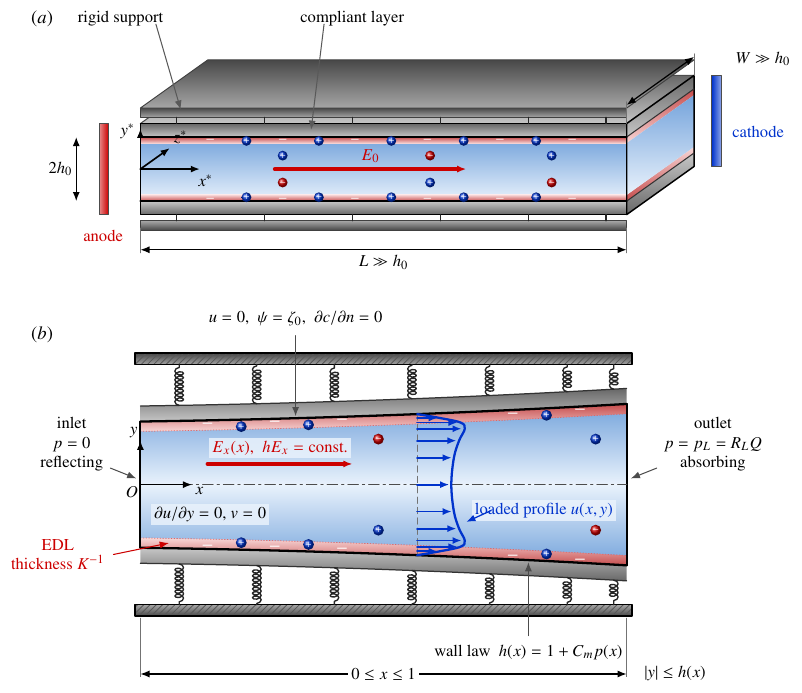}
		\caption{Electroosmotic pump with compliant slit walls. (a) Dimensional geometry: half-height $h_0$, length $L$, width $W$ and compliant-layer thickness $h_c$. The reference electric field is $E_0$. (b) Reduced domain, $0\leq x\leq1$ and $|y|\leq h(x)$, with $p(0)=0$ and $p_L=R_LQ$. The solute inlet is reflecting and the outlet absorbing.}
		\label{fig:schematic}
	\end{figure}
	
	\subsection{Geometry, scales and electrohydrodynamic equations}
	\label{sec:geometry}
	
	We consider the configuration shown in figure~\ref{fig:schematic}. A symmetric monovalent electrolyte, with $z_+=1$ and $z_-=-1$, occupies a long, shallow slit microchannel. The axial coordinate is $x^*$, and the transverse coordinate $y^*$ is measured from the channel centreline. The channel width is much larger than the gap, so sidewall effects and spanwise variations of the flow, wall deformation and electric current are neglected. Flow rate and electric current are therefore reported per unit span. The undeformed channel has half-height $h_0$, length $L\gg h_0$ and compliant layers of thickness $h_c$ bonded to rigid exterior supports. The walls deform symmetrically and are located at $y^*=\pm h^*(x^*)$, where $h^*(x^*)=h_0+h_1^*(x^*)$ and $h_1^*$ is the outward displacement of either wall. The geometric parameters are the aspect ratio $\delta=h_0/L\ll1$, wall-thickness ratio $\beta=h_c/h_0$, foundation slenderness $h_c/L=\beta\delta\ll1$ and deformation amplitude $\chi=\max_{x^*}|h_1^*|/h_0$. The electrolyte has density $\rho_f$, dielectric permittivity $\varepsilon_e$, uniform bulk conductivity $\sigma$ and zero-shear viscosity $\eta_0$. A positive reference electric field is defined by $E_0=\Delta\Phi_{\mathrm{app}}^*/L$ under constant-voltage operation and by $E_0=I_{\mathrm{app}}^*/(2\sigma h_0)$ under constant-current operation, where $I_{\mathrm{app}}^*$ is the imposed current per unit span. The two operating modes are compared at the same undeformed reference field. For the negative reference wall potential considered here, $\zeta_0<0$, the positive Helmholtz--Smoluchowski velocity scale is $U_{\mathrm{hs}}=-\varepsilon_e\zeta_0E_0/\eta_0$ \citep{Ghosal2006Electrokinetic}. Axial distance is scaled by $L$, transverse distance and wall displacement by $h_0$, axial velocity by $U_{\mathrm{hs}}$, transverse velocity by $\delta U_{\mathrm{hs}}$, pressure by $p_0=\eta_0U_{\mathrm{hs}}L/h_0^2$, extra stress by $\eta_0U_{\mathrm{hs}}/h_0$, time by $L/U_{\mathrm{hs}}$ and flow rate by $U_{\mathrm{hs}}h_0$. Thus,
	$x=x^*/L$, $y=y^*/h_0$, $h=h^*/h_0$, $u=u^*/U_{\mathrm{hs}}$, $v=v^*/(\delta U_{\mathrm{hs}})$, $p=p^*/p_0$, $\boldsymbol{\tau}=\boldsymbol{\tau}^*/(\eta_0U_{\mathrm{hs}}/h_0)$, $t=U_{\mathrm{hs}}t^*/L$, $Q=Q^*/(U_{\mathrm{hs}}h_0)$ and $c=c^*/c_0$, where $c_0$ is an arbitrary concentration scale. The electrical variables are $\phi=\phi^*/(E_0L)$, $\Psi=\psi^*/\zeta_0$, $E_x=E_x^*/E_0$ and $\rho_e=\rho_e^*/(-\varepsilon_e\zeta_0/h_0^2)$.
	
	The flow is assumed to be steady, incompressible and inertia-free. At local ionic equilibrium, the electrical force associated with the EDL potential is balanced by the corresponding ionic osmotic-stress gradient. The remaining mechanical balance is
	\begin{align}
		\boldsymbol{\nabla}^{*}\boldsymbol{\cdot}\boldsymbol{u}^*&=0,
		\label{eq:continuity}\\
		\boldsymbol{0}
		&=-\boldsymbol{\nabla}^{*}p^*
		+\boldsymbol{\nabla}^{*}\boldsymbol{\cdot}\boldsymbol{\tau}^*
		+\rho_e^*E_x^*(x^*)\boldsymbol{e}_x,
		\label{eq:momentum}
	\end{align}
	where $p^*$ is the mechanical gauge pressure relative to the exterior. The reference Reynolds number is $\mathrm{Re}=\rho_fU_{\mathrm{hs}}h_0/\eta_0\ll1$, and the loaded condition $\mathrm{Re}\max_{x,y}(|u|h)\ll1$ is checked a posteriori.
	
	The total potential is decomposed as $\Phi^*(x^*,y^*)=\phi^*(x^*)+\psi^*(x^*,y^*)$, where $\phi^*$ is the slowly varying axial conduction potential and $\psi^*$ is the equilibrium EDL potential. The remaining axial electrical forcing in \eqref{eq:momentum} is due to the conduction field $E_x^*=-\mathrm{d}\phi^*/\mathrm{d}x^*$. Under the Debye--H\"uckel approximation \citep{Ghosal2006Electrokinetic},
	\begin{equation}
		\boldsymbol{\nabla}^{*}\boldsymbol{\cdot}
		\left(\varepsilon_e\boldsymbol{\nabla}^{*}\psi^*\right)
		=-\rho_e^*,
		\qquad
		\rho_e^*=-\varepsilon_e\kappa^2\psi^*,
		\label{eq:poisson}
	\end{equation}
	where $\kappa^{-1}$ is the Debye length and $K=\kappa h_0$ is the Debye parameter. The approximation requires $|e\zeta_0|/(k_{\mathrm B}T)\ll1$. The boundary conditions are $\psi^*(x^*,\pm h^*)=\zeta_0$ and $\partial\psi^*/\partial y^*=0$ at $y^*=0$. In pressure-free flow, a plug-like profile requires the local condition $Kh(x)\gg1$.
	
	Within the bulk-Ohmic approximation, $i_x^*=\sigma E_x^*$ and current conservation gives $I^*=2\sigma h^*E_x^*=\text{constant}$ \citep{Ghosal2002Lubrication,Ghosal2006Electrokinetic}. This closure assumes uniform conductivity, insulating walls, negligible Faradaic current and concentration polarisation, and surface and streaming currents small relative to the bulk conduction current. Maxwell and equilibrium ionic-osmotic contributions to the wall-normal traction are neglected at leading order. Their maximum magnitude relative to $p_0$ is denoted by $\mathcal{T}_{\mathrm{el}}$; the conditions $\mathcal{T}_{\mathrm{el}}\ll1$ and $C_m\mathcal{T}_{\mathrm{el}}\ll1$ ensure that both the neglected traction and its associated wall displacement remain small.
	
	\subsection{Constitutive model and lubrication assumptions}
	\label{sec:constitutive}
	
	The steady extra stress is described by the linear form of the simplified Phan-Thien--Tanner model \citep{PhanThien1977New,Afonso2009Analytical}. With $(\boldsymbol{\nabla}^{*}\boldsymbol{u}^*)_{ij}=\partial u_i^*/\partial x_j^*$,
	\begin{equation}
		f \left(\operatorname{tr}\boldsymbol{\tau}^*\right)
		\boldsymbol{\tau}^*
		+\lambda\left[
		\boldsymbol{u}^*\boldsymbol{\cdot}
		\boldsymbol{\nabla}^{*}\boldsymbol{\tau}^*
		-\left(\boldsymbol{\nabla}^{*}\boldsymbol{u}^*\right)
		\boldsymbol{\cdot}\boldsymbol{\tau}^*
		-\boldsymbol{\tau}^*
		\boldsymbol{\cdot}
		\left(\boldsymbol{\nabla}^{*}\boldsymbol{u}^*\right)^T
		\right]
		=2\eta_p\boldsymbol{D}^*,
		\label{eq:sptt}
	\end{equation}
	where $\boldsymbol{D}^*=[\boldsymbol{\nabla}^{*}\boldsymbol{u}^*+(\boldsymbol{\nabla}^{*}\boldsymbol{u}^*)^T]/2$ and
	$f=1+\epsilon_p(\lambda/\eta_p)\operatorname{tr}\boldsymbol{\tau}^*$. Here, $\eta_p$, $\lambda$ and $\epsilon_p$ are the polymer viscosity, relaxation time and extensibility parameter. We consider the zero-Newtonian-solvent-viscosity limit, $\eta_s=0$ and $\eta_0=\eta_p$, so $\boldsymbol{\tau}^*$ is the total extra stress.
	
	At leading lubrication order, the flow is locally unidirectional and dominated by transverse shear \citep{Ghosal2002Lubrication}. Under the assumed slow axial variation, material advection in \eqref{eq:sptt} is $O(\delta)$ relative to the local deformation terms. In the axial momentum balance, the characteristic axial normal-stress gradient is $O(\delta \mathrm{Wi}_{\mathrm{loc}})$ relative to the characteristic transverse shear-stress gradient, where $\mathrm{Wi}_{\mathrm{loc}}=\lambda\max_{x^*,y^*}|\partial u^*/\partial y^*|$. The reduction therefore requires both $\delta\ll1$ and $\delta \mathrm{Wi}_{\mathrm{loc}}\ll1$. Analyses of slowly varying Oldroyd--B channels \citep{Boyko2022Pressure} and FENE--CR contractions \citep{Mahapatra2025Finite} resolve elastic-stress contributions to the pressure drop. Here the stated smallness condition controls the omission of the axial normal-stress gradient in the local shear reduction. The local simple-shear balances give $\tau_{yy}^*=\tau_{zz}^*=0$, $\tau_{xx}^*=2(\lambda/\eta_p)(\tau_{xy}^*)^2$ and $\operatorname{tr}\boldsymbol{\tau}^*=\tau_{xx}^*$, therefore
	\begin{equation}
		\frac{\partial u^*}{\partial y^*}
		=\frac{\tau_{xy}^*}{\eta_p}
		+2\epsilon_p\frac{\lambda^2}{\eta_p^3}
		\left(\tau_{xy}^*\right)^3.
		\label{eq:shear}
	\end{equation}
	The cubic dependence on the total shear stress prevents linear superposition of the electroosmotic and pressure-driven velocity contributions. The electroosmotic shear rate is maximum within the EDL. We define $\mathrm{De}_{\kappa}=\lambda\kappa U_{\mathrm{hs}}$ and the height-based parameter $\mathrm{Wi}_h=\lambda U_{\mathrm{hs}}/h_0=\mathrm{De}_{\kappa}/K$. The notation $\mathrm{De}_{\kappa}$ is retained for consistency with the electroosmotic-rheology convention used below, although it is based on a shear-rate scale. The leading rheological correction enters through $g=\epsilon_p\mathrm{De}_{\kappa}^2$, but the asymptotic validity condition involving $\mathrm{Wi}_{\mathrm{loc}}$ cannot be expressed in terms of $g$ alone.
	
	\subsection{Passive-solute transport and validity conditions}
	\label{sec:solute}
	
	A dilute neutral solute with constant diffusivity $D_s^*$ is transported passively by the steady flow. The constant-diffusivity description is intended for sufficiently small molecular tracers, for which diffusion is approximated by its solvent value. For colloidal tracers, rheology-dependent diffusivity modifies the Taylor--Aris dispersion, as discussed for Carreau and Phan-Thien--Tanner flows by \citet{Boyko2026Taylor}. We assume that the solute has no electrophoretic velocity, does not adsorb onto or react with the channel walls, and does not alter the viscosity, electrical conductivity, or velocity field. Its concentration therefore satisfies the following advection–diffusion equation:
	\begin{equation}
		\frac{\partial c^*}{\partial t^*}
		+\boldsymbol{\nabla}^{*}\boldsymbol{\cdot}
		\left(\boldsymbol{u}^*c^*\right)
		=D_s^*\nabla^{*2}c^*.
		\label{eq:solute}
	\end{equation}
	The gap-based P\'eclet number is defined as $\mathrm{Pe}=U_{\mathrm{hs}}h_0/D_s^*$. At the stationary walls, $\boldsymbol{u}^*=\boldsymbol{0}$ and $\boldsymbol{n}\boldsymbol{\cdot}\boldsymbol{\nabla}^{*}c^*=0$, where $\boldsymbol{n}$ denotes the outward unit normal. Axial inlet and outlet conditions are specified with the reduced first-passage problem in Section~\ref{sec:reduction}. The Taylor--Aris reduction \citep{Taylor1953Dispersion,Aris1956} applies after transverse diffusion has equilibrated the solute across the gap. On the reference scales, this requires $\mathrm{Pe} \delta\ll1$. For loaded states, the corresponding loaded-state condition $\mu_{\max}=\mathrm{Pe} \delta\max_x[U_{\mathrm{loc}}(x)h^2(x)]\ll1$ is checked a posteriori, where $U_{\mathrm{loc}}(x)=\max_y|u(x,y)|$.
	
	\subsection{Wall mechanics and electrical forcing}
	\label{sec:wall_electrical}
	
	The walls are modelled as a compressible elastic layer bonded to a rigid exterior support. Under 1D constrained compression, the normal stress and strain satisfy $p^*=(2G+\lambda_s)h_1^*/h_c$, where $G$ and $\lambda_s$ are the Lam\'e parameters \citep{Skotheim2005Soft,Chandler2020Validity}. The dimensional and dimensionless wall laws are therefore
	\begin{equation}
		h_1^*(x^*)=\frac{h_c}{2G+\lambda_s} p^*(x^*),
		\qquad
		h(x)=1+C_mp(x),
		\qquad
		C_m=\frac{h_c}{h_0}
		\frac{\eta_0U_{\mathrm{hs}}L}
		{h_0^2(2G+\lambda_s)}.
		\label{eq:winkler}
	\end{equation}
	For pressure variations on the scale $L$, the local foundation law assumes $(h_c/L)^2\ll 1-2\nu_s$, where $\nu_s$ is the Poisson ratio of the coating \citep{Chandler2020Validity}. Positive internal pressure increases the channel half-height. The wall law remains linear, but the full dependence of the flow equations on $h=1+C_mp$ is retained without expansion in $C_mp$. The maximum dimensionless displacement is $\chi=C_m\max_x|p|$, while the maximum constrained wall strain is $\varepsilon_{\mathrm{w},\max}=\chi/\beta=C_m\max_x|p|/\beta$. The local-foundation and lubrication descriptions require $\beta\delta\ll1$, $\chi\ll1$, $\varepsilon_{\mathrm{w},\max}\ll1$, $h>0$ and $\delta\max_x|h'|\ll1$, where a prime denotes differentiation with respect to $x$. The wall response is assumed purely elastic and time independent. Dimensionless current conservation gives $hE_x=\text{constant}$. With the reference-field definitions above,
	\begin{equation}
		E_x(x)=
		\begin{cases}
			h^{-1}(x),
			& \text{constant current},\\[2mm]
			\displaystyle
			\frac{h^{-1}(x)}
			{\int_0^1h^{-1}(s) \mathrm{d}s},
			& \text{constant voltage}.
		\end{cases}
		\label{eq:field_closures}
	\end{equation}
	Deformation therefore modulates the local field under both operating modes. Constant-current operation fixes the normalisation directly, whereas the constant-voltage normalisation is determined non-locally by the complete channel shape. Table~\ref{tab:regime} summarises the small-parameter and a posteriori checks for the reference case; the corresponding validity conditions are stated above.
	
	\subsection{Operating configuration and performance measures}
	\label{sec:operating}
	
	The channel operates as an electroosmotic pump delivering fluid in the positive $x$-direction against a downstream hydraulic load. The inlet is maintained at ambient pressure, $p(0)=0$, and the positive outlet pressure $p_L=p(1)$ is a back pressure; hence a positive axial pressure gradient is adverse to EOF.
	
	The pump characteristic is parameterised by prescribing the flow rate $Q$. For each $Q$, the mixed-flow relation derived in Section~\ref{sec:reduction} is inverted locally for the pressure gradient and integrated from the inlet to obtain $p_L(Q)$. A downstream load law $p_L=\mathcal{F}_L(Q)$ selects an operating point. For the linear resistance illustrated in figure~\ref{fig:schematic}, $p_L=R_LQ$, where $R_L=R_L^*h_0^3/(\eta_0L)$. Prescribed throughput, prescribed back pressure and prescribed hydraulic resistance are therefore distinct operating specifications. Free flow is defined by $p_L=0$ and $Q=Q_{\mathrm{free}}$, with $Q_{\mathrm{free}}$ evaluated at the same imposed current or voltage, Debye parameter and rheological properties as the loaded state. For the uniform reference channel, $p=0$, $h=1$ and $E_x=1$ constitute the pressure-free solution. Stall is defined by $Q=0$ and $p_L=p_{\mathrm{stall}}$ and is approached as $R_L\rightarrow\infty$ for the linear load. Forward-pumping states between free flow and stall are parameterised by $r=Q/Q_{\mathrm{free}}$, with $0\leq r\leq1$. Fixing $r$ while changing $K$ or the rheological parameters does not generally fix the absolute flow rate.
	
	The distributed pressure loading of the compliant walls is measured by $\mathcal{W}=\int_0^1p(x) \mathrm{d}x$. The dimensionless channel length is unity, so $\mathcal{W}$ is also the mean dimensionless pressure and is distinct from the stall head. Deformation-induced field modulation shifts the pump characteristic and hence the operating-point intersection with a prescribed downstream load.
	
	\section{Reduced order flow and Taylor--Aris dispersion}\label{sec:reduction}
	
	This section develops the reduced description of the coupled electroosmotic flow and passive-solute dispersion. Pump characteristics are obtained under constant-voltage (CV) and constant-current (CC) operation, whereas detailed dispersion calculations are reported for CC operation. Within the retained leading-order flow equations, the extensibility parameter and Debye-length-based viscoelastic parameter enter through $g=\epsilon_p\mathrm{De}_{\kappa}^{2}$; the validity condition involving $\mathrm{Wi}_{\mathrm{loc}}$ does not collapse in $g$ alone. The $O(\chi)$ expansion in Appendix~\ref{app:field_expansions} is used only to interpret the first-order electric-field modulation. All loaded pump and transport calculations retain the unexpanded relations $h=1+C_mp$ and \eqref{eq:field_closures}, subject to the validity conditions stated in Section~\ref{sec:formulation}. All variables below are dimensionless, and asterisks are omitted.
	
	\subsection{Reduced equations and electrostatic potential}
	\label{sec:reduced_electrostatics}
	
	In the lubrication limit, the leading-order momentum and continuity equations are
	\begin{align}
		\partial_xu+\partial_yv&=0, \\
		p'(x)&=\partial_y\tau_{xy}+\rho_eE_x(x), \\
		\partial_yp&=0,
	\end{align}
	where $p'=\mathrm{d}p/\mathrm{d}x$ and $p'>0$ is adverse to flow in the positive $x$-direction. Hence, $p=p(x)$ is uniform across the local gap. The leading transverse electrostatic problem is
	\begin{equation}
		\partial_{yy}\Psi=K^2\Psi,
		\qquad
		\Psi(x,\pm h(x))=1,
		\qquad
		\Psi(x,y)
		=\frac{\cosh(Ky)}{\cosh\bigl(Kh(x)\bigr)},
		\label{eq:local_EDL}
	\end{equation}
	and the corresponding dimensionless charge density is
	\begin{equation}
		\rho_e(x,y)=K^2\Psi(x,y).
		\label{eq:local_charge}
	\end{equation}
	These expressions apply for arbitrary local $Kh$ within the regime conditions in Section~\ref{sec:formulation}. The loaded calculations use the full electrical closures \eqref{eq:field_closures}; the small-deformation expansions in Appendix~\ref{app:field_expansions} show how the global voltage normalisation removes the mean-deformation contribution under CV operation.
	
	\subsection{Undeformed pressure-free reference state}
	\label{sec:reference_state}
	
	Consider the undeformed, pressure-free reference state, denoted by the superscript $(0,0)$, for which $h=1$, $p'=0$ and $E_x=1$. Integration of the sPTT shear relation with no slip gives the velocity profile in Appendix~\ref{app:reference_state} and the flow rate per unit span
	\begin{equation}
		Q^{(0,0)}
		=2\left(1-\frac{\tanh K}{K}\right)
		+g \mathcal{Q}_1(K),
		\label{eq:Q00}
	\end{equation}
	where $\mathcal Q_1(K)$ is given by \eqref{eq:Q1}. Along the fixed-$g$ thin-EDL path, $\mathcal{Q}_1\to4/3$ and therefore $Q^{(0,0)}\to2+4g/3$ as $K\to\infty$. The limiting correction represents a finite viscoelastic contribution to the effective electroosmotic slip.
	
	\subsection{Mixed electroosmotic and pressure-driven flow rate}
	\label{sec:mixed_flux}
	
	Under simultaneous electroosmotic and pressure-driven forcing, the total shear stress follows from the axial momentum balance and centreline symmetry:
	\begin{equation}
		\tau_{xy}(x,y)
		=p'(x)y
		-E_x(x)K
		\frac{\sinh(Ky)}{\cosh \bigl(Kh(x)\bigr)},
		\qquad
		\tau_{xy}(x,0)=0.
		\label{eq:taumixed}
	\end{equation}
	Writing $a=2g/K^2$, integration of $\partial_yu=\tau_{xy}+a\tau_{xy}^3$ with no slip at $y=h$ gives
	\begin{equation}
		Q=-2\int_0^h
		y\left(\tau_{xy}+a\tau_{xy}^3\right) \mathrm{d}y.
		\label{eq:Qexact}
	\end{equation}
	The Newtonian contribution is \citep{Dutta2001Combined}
	\begin{equation}
		Q_{\mathrm N}
		=2E_x
		\left[
		h-\frac{\tanh(Kh)}{K}
		\right]
		-\frac{2}{3}h^3p',
		\label{eq:QN_mixed}
	\end{equation}
	whereas the cubic correction contains terms proportional to $p'^3$, $p'^2E_x$, $p'E_x^2$ and $E_x^3$. The first and last are the pure pressure-driven and pure electroosmotic viscoelastic contributions, respectively, while the middle two are mixed contributions omitted by a superposed model \citep{Afonso2009Analytical}. The complete closed-form expression is given in \S S3 of the supplementary material. At fixed $h$ and $E_x$, differentiation of \eqref{eq:Qexact} gives
	\begin{equation}
		\left.
		\frac{\partial Q}{\partial p'}
		\right|_{h,E_x}
		=-2\int_0^h
		y^2\left(1+3a\tau_{xy}^2\right) \mathrm{d}y<0,
		\qquad
		h>0,\quad a\geq0.
		\label{eq:monotone}
	\end{equation}
	Thus, the mixed-flow relation has a unique local pressure-gradient root at each axial station during the prescribed-throughput calculation. Under CV operation, this local inversion is coupled to the non-local voltage constraint through \eqref{eq:field_closures}. For fixed $Q$ and conserved $hE_x$, the local inversion and the wall law give a smooth autonomous equation for $p(x)$. If $p(0)=p(1)=0$, then $p'=0$ at an interior point, and uniqueness implies $p\equiv0$. Thus $h=1$ and both electrical closures give $E_x=1$, so $Q_{\mathrm{free}}=Q^{(0,0)}$.
	
	\subsection{Axial solute dispersion}
	\label{sec:axial_dispersion}
	
	The transport model uses the loaded velocity field determined by \eqref{eq:taumixed} and the sPTT shear relation. For a local quantity $f(x,y)$, define the cross-sectional average
	\begin{equation}
		\langle f\rangle_y
		=\frac{1}{2h(x)}
		\int_{-h(x)}^{h(x)}f(x,y) \mathrm{d}y.
		\label{eq:cross_sectional_average}
	\end{equation}
	The mean velocity and concentration are $\bar u=\langle u\rangle_y=Q/(2h)$ and $\bar c=\langle c\rangle_y$, respectively. At each axial station, the Taylor--Aris cell problem \citep{Aris1956} is
	\begin{equation}
		-\partial_{yy}\varphi=u-\bar u,
		\qquad
		\partial_y\varphi(x,\pm h)=0,
		\qquad
		\langle\varphi\rangle_y=0,
		\label{eq:cellloaded}
	\end{equation}
	and the local shear-dispersion coefficient is
	\begin{equation}
		\mathcal{K}(x)
		=\left\langle
		\left(\partial_y\varphi\right)^2
		\right\rangle_y
		\geq0.
		\label{eq:local_dispersion}
	\end{equation}
	For a rigid uniform channel, \eqref{eq:cellloaded} reduces to the classical Taylor--Aris cell problem based on the corresponding fully developed velocity profile. For a slowly varying gap with local area per unit span $A(x)=2h(x)$, conservation of solute gives the cross-sectionally averaged transport equation \citep{Datta2008Dispersion}
	\begin{equation}
		\frac{\partial}{\partial t}\left(A\bar c\right)
		+\frac{\partial}{\partial x}\left(Q\bar c\right)
		=
		\delta\frac{\partial}{\partial x}
		\left[
		A\left(
		\mathrm{Pe}^{-1}
		+\mathrm{Pe} \mathcal{K}(x)
		\right)
		\frac{\partial\bar c}{\partial x}
		\right].
		\label{eq:transport}
	\end{equation}
	The reduction is derived in the mapped coordinate $\eta=y/h(x)$ in \S S8 of the supplementary material, where the transverse velocity is retained. The neglected flux is $O \left(\mu^2+\delta|h'|\right)$ relative to the retained contribution, with $\mu=\mathrm{Pe} \delta$, and the loaded validity condition is given in Section~\ref{sec:formulation}. Writing the one-dimensional solute density as $P(x,t)=A(x)\bar c(x,t)$, \eqref{eq:transport} corresponds to a diffusion process with
	\begin{equation}
		D_{\mathrm{ax}}(x)
		=\delta\left(
		\mathrm{Pe}^{-1}
		+\mathrm{Pe} \mathcal{K}(x)
		\right),
		\qquad
		b(x)
		=\bar u+D_{\mathrm{ax}}'
		+D_{\mathrm{ax}}\frac{A'}{A},
		\label{eq:effective_drift}
	\end{equation}
	where primes denote derivatives with respect to $x$. Let $T(x)$ be the mean first-passage time to the outlet for a particle starting at $x$, and let $V(x)$ be the corresponding centred second moment. The backward first-passage formulation \citep{Gardiner2009Stochastic} gives
	\begin{equation}
		bT'+D_{\mathrm{ax}}T''=-1,
		\qquad
		bV'+D_{\mathrm{ax}}V''
		=-2D_{\mathrm{ax}}(T')^2,
		\label{eq:backward}
	\end{equation}
	with reflecting conditions $T'(0)=V'(0)=0$ at the inlet and absorbing conditions $T(1)=V(1)=0$ at the outlet. The mean transit time, temporal variance and plate number are
	\begin{equation}
		t_R=T(0),
		\qquad
		\sigma_t^2=V(0),
		\qquad
		N=\frac{t_R^2}{\sigma_t^2}.
		\label{eq:plate_number}
	\end{equation}
	
	Molecular diffusion is retained, so the first-passage moments remain finite as $Q\to0$ under the chosen reflecting--absorbing boundary conditions. At stall, outlet passage occurs through effective axial diffusion, including molecular and shear-enhanced contributions. This limit is mathematically well defined but is not interpreted as an ordinary finite-throughput separation state.
	
	In the advection-dominated regime, the first-passage moments reduce to the axial integrals in Appendix~\ref{app:transit_moments}. The variance weights local spreading by $\bar u^{-3}$, so slow portions of the channel contribute disproportionately to band broadening. This approximation is not used near stall, where $\bar u\to0$. Detailed transport maps are reported for CC operation. CV operation changes the local field normalisation and therefore shifts the quantitative transport predictions. At fixed $K$ and fixed local $h$, $E_x$ and $p'$, the velocity and cell function are affine in $g$, giving the quadratic dispersion coefficient derived in Appendix~\ref{app:local_rheology}. Along a prescribed-throughput characteristic, $p'$, $h$ and $E_x$ also depend on $g$, so the loaded coefficient need not be quadratic.
	
	\subsubsection{Thin-EDL plateau and its physical origin}
	\label{sec:thin_EDL_plateau}
	
	The thin-EDL limit separates shear confined to the charged layers from shear sustained across the channel by the hydraulic load. We take $K\to\infty$ at fixed $g$, with $h$, $E_x$ and $p'$ bounded and $h>0$. The coefficient $a=2g/K^2$ then tends to zero, so the cubic shear-rate contribution vanishes in the core. Within each EDL, however, the $O(K)$ stress produces a finite viscoelastic contribution to the effective slip:
	\begin{equation}
		U_s(x)=E_x(x)+\frac{2}{3}gE_x^3(x).
		\label{eq:effective_slip}
	\end{equation}
	Matching the EDL solution to the core gives, at leading order,
	\begin{equation}
		u(x,y)
		=U_s(x)
		+\frac{p'(x)}{2}
		\left[y^2-h^2(x)\right],
		\qquad
		\bar u(x)
		=U_s(x)-\frac{p'(x)h^2(x)}{3}.
		\label{eq:thin_EDL_velocity}
	\end{equation}
	Applying the cell problem to this core profile gives the local limiting coefficient
	\begin{equation}
		\mathcal{K}(x)
		\longrightarrow
		\frac{2}{105}
		h^2(x)
		\left[
		U_s(x)-\bar u(x)
		\right]^2,
		\qquad
		K\to\infty
		\quad\text{at fixed }g.
		\label{eq:anchor}
	\end{equation}
	The pressure-driven velocity deficit $U_s-\bar u=p'h^2/3$ is therefore the amplitude that controls the plateau. The cell response is linear in this deficit, while its contribution to dispersion is quadratic. Thinning the EDL confines the electroosmotic velocity deficit to a progressively smaller fraction of the gap; the pressure-driven core shear needed to sustain a finite load remains. The cell integration yielding the factor $2/105$ is given in \S S7 of the supplementary material. For a rigid uniform channel, $Q_{\mathrm{free}}\to2hU_s$ and $\bar u/U_s\to r$, so \eqref{eq:anchor} reduces to
	\begin{equation}
		\mathcal{K}
		\longrightarrow
		\frac{2}{105}
		U_s^2h^2(1-r)^2.
		\label{eq:rigid_thin_EDL_dispersion}
	\end{equation}
	For the case of rigid reference slit, $h=E_x=1$ and $S=1+2g/3$. The adverse pressure gradient reduces the mean velocity by $S(1-r)$, giving a dispersion plateau of $2S^2(1-r)^2/105$. At a fixed $r$, increasing $g$ raises the plateau through the effective slip $S$. The limiting centreline velocity is $u(0)\to S(3r-1)/2$, so the plateau remains positive for $1/3<r<1$, where the core flow remains forward.
	
	\subsubsection{Compliance and electrical operation}
	\label{sec:thin_EDL_compliance}
	
	In a compliant microchannel, the global throughput fraction generally differs from the local ratio $\bar u/U_s$. Write $E_x=\mathcal J/h$, where $\mathcal J=1$ under CC operation and $\mathcal J=\langle h^{-1}\rangle_x^{-1}$ under CV operation, with $\langle f\rangle_x=\int_0^1f(x) \mathrm{d}x$. Since the free-flow reference state is undeformed in both modes, $Q_{\mathrm{free}}\to2S$, and \eqref{eq:anchor} becomes
	\begin{equation}
		\mathcal K_\infty(x)
		=\frac{2}{105}
		\left[\mathcal J+\frac{2g\mathcal J^3}{3h^2(x)}-rS\right]^2.
		\label{eq:thin_EDL_compliant_plateau}
	\end{equation}
	Here, the terms $h$ and $\mathcal J$ are evaluated in the self-consistent limiting pump state. Under the CC operation, a Newtonian fluid has $hU_s=1$ and $h\bar u=r$, so the gap dependence cancels exactly from this leading plateau: $\mathcal K_\infty=2(1-r)^2/105$. Compliance still changes the pressure and transit time. For $g>0$, wall expansion weakens the term $2g/(3h^2)$; on the branch with $p'>0$, it lowers the local plateau relative to the rigid reference at the same $r$ and $g$.
	
	Under CV operation, the global field normalisation also changes, so the CC cancellation and compliance trend do not generally apply under CV operation. For example, for a Newtonian fluid $\mathcal K_\infty=2(\mathcal J-r)^2/105$, which exceeds its rigid value when positive pressure expands the channel and hence $\mathcal J>1$. More generally, the weak-compliance expansion in \S S7 gives relative changes of $-4gC_m$ under CC operation and $(3+2g)C_m$ under CV operation for the path-averaged plateau, to first order at fixed $r<1$ and $g$. These limiting statements concern the shear-dispersion coefficient; neither its finite-$K$ trend nor the plate number follows from the plateau alone.
	
	\subsubsection{Crossover from free flow and endpoint limits}
	\label{sec:thin_EDL_crossover}
	
	For the rigid reference slit, put $B=1+2g/9$. The integrated velocity deficit of one EDL is $B/K$, giving $\mathcal K_{\mathrm{free}}\sim B^2/(3K^2)$. Balancing this contribution against the loaded plateau defines the crossover estimate
	\begin{equation}
		K_c=\sqrt{\frac{35}{2}} \frac{B}{S(1-r)}.
		\label{eq:thin_EDL_crossover_scale}
	\end{equation}
	For weak loading, $K_c\gg1$: the EDL contribution to dispersion dominates for $1\ll K\ll K_c$, the two contributions interact for $K=O(K_c)$, and the core-shear plateau dominates for $K\gg K_c$. In the joint limit $K\to\infty$, $1-r=O(K^{-1})$, the cell calculation in \S S7 gives
	\begin{equation}
		\mathcal K
		=\frac{B^2}{3K^2}
		-\frac{2SB(1-r)}{15K}
		+\frac{2S^2(1-r)^2}{105}
		+O(K^{-3}).
		\label{eq:thin_EDL_weak_load}
	\end{equation}
	The negative cross term expresses cancellation between the EDL and pressure-driven contributions to the cell response and explains the interior dispersion minimum. Its leading location is $K_{\min}\sim5B/[S(1-r)]$ as $r\to1$; it differs from the balance scale $K_c$. The plate-number optimum also depends on molecular diffusion and transit-time weighting. Equation~\eqref{eq:thin_EDL_weak_load} is a weak-loading asymptotic result, not a finite-$K$ approximation for arbitrary load. A compliant channel additionally requires the spatially varying fields in \eqref{eq:thin_EDL_compliant_plateau}.
	
	As $r\to1$, the rigid-channel plateau vanishes quadratically and $K_c$ diverges as $(1-r)^{-1}$. Taking $K\to\infty$ and $r\to1$ in either order gives zero for the unscaled coefficient, but its leading behaviour depends on $K(1-r)$; the recovery of free-flow scaling is therefore non-uniform in $K$. At the opposite endpoint, $r\to0$, the rigid thin-EDL coefficient tends to $2S^2/105$ despite zero mean flow. The opposing motions still generate shear-enhanced spreading. Outlet passage under the reflecting--absorbing boundary conditions is then diffusion dominated, and a finite first-passage plate number does not signify finite-throughput separation. Molecular diffusion remains present at both endpoints.
	
	\section{Numerical simulation and model validation}\label{sec:num}
	
	The numerical checks distinguish the evaluation of the reduced flow relations, the accuracy of the Taylor--Aris transport reduction and the convergence of the discretised calculations. For each prescribed flow rate, the local mixed-flow relation is inverted to obtain the pressure gradient, and the pressure is advanced subject to $h=1+C_mp$. The cell problem is evaluated by quadrature and the backward moment equations are solved on the resulting axial fields. Direct integration and analytical limiting cases verify the reduced equations and their implementation; Brownian-dynamics simulations retaining the transverse particle motion test the subsequent transport reduction in the same flow field. These comparisons do not independently validate the constitutive, electrostatic or wall-mechanical assumptions, whose applicability is assessed through the regime conditions in Section~\ref{sec:regime_checks}. Unless stated otherwise, all variables are dimensionless and follow Sections~\ref{sec:formulation} and \ref{sec:reduction}.
	
	\subsection{Verification of the flow relations}
	\label{sec:flow_validation}
	
\begin{figure}
		\centering
		\includegraphics[width=0.49\textwidth]{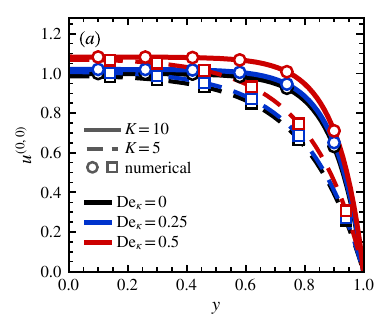}
		\hfill
		\includegraphics[width=0.49\textwidth]{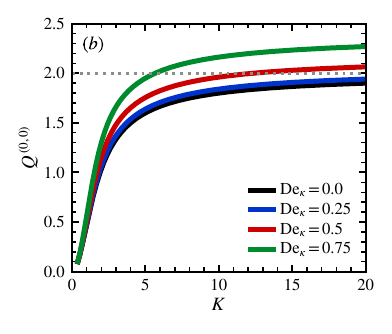}
		\caption{Rigid-channel electroosmotic flow at $\epsilon_p=0.5$. (a) Velocity profiles for $K=5,10$ and $\mathrm{De}_{\kappa}=0,0.25,0.5$: lines, \eqref{eq:u00}; symbols, direct integration. (b) Flow rate versus $K$ for $\mathrm{De}_{\kappa}=0,0.25,0.5,0.75$. Dotted line: Newtonian thin-EDL limit.}
		\label{fig:flow}
	\end{figure}

	The pressure-free rigid-channel profile \eqref{eq:u00} is first checked by direct numerical integration of the shear relation subject to the no-slip condition. In figure~\ref{fig:flow}(a), the closed-form and numerical profiles agree at the plotted resolution for $K=5,10$ and $\mathrm{De}_{\kappa}=0,0.25,0.5$, with $\epsilon_p=0.5$. This comparison tests the integration of the constitutive shear relation and enforcement of the wall condition in the undeformed reference state. Figure~\ref{fig:flow}(b) shows the corresponding flow rate \eqref{eq:Q00} and its Newtonian thin-EDL limit. The nonlinear mixed terms that arise under load require the separate checks below.
	
	The closed-form evaluation of the mixed-flow integral \eqref{eq:Qexact}, given in \S S3 of the supplementary material, is compared with direct quadrature over 600 combinations of $K$, $h$, $E_x$, $p'$ and $g$ spanning the computational range. The dimensionless fluxes agree to within $3.2\times10^{-11}$. The Newtonian electroosmotic and pressure-driven contributions, $2E_x[h-\tanh(Kh)/K]$ and $-2h^3p'/3$, are recovered to machine precision, as is \eqref{eq:Q00} at $p'=0$, $h=E_x=1$. These are algebraic consistency checks, since both evaluations use the same reduced stress relation. The computed derivative $\partial Q/\partial p'$ remains negative throughout the sampled states, with largest value $-0.486$, consistent with the analytical monotonicity result \eqref{eq:monotone}; the global sign follows from that analytical result. The error introduced by omitting nonlinear mixed terms is assessed separately using
	\begin{align}
		Q_{\mathrm{sup,N}}
		&=Q_{\mathrm{EOF}}^{\mathrm{sPTT}}
		-\frac{2}{3}h^3p',
		\label{eq:QsupN}\\
		Q_{\mathrm{sup,sPTT}}
		&=Q_{\mathrm{EOF}}^{\mathrm{sPTT}}
		-\frac{2}{3}h^3p'
		-\frac{4gh^5}{5K^2}p'^3.
		\label{eq:QsupsPTT}
	\end{align}
	The first approximation treats the pressure-driven component as Newtonian, whereas the second retains the pure sPTT pressure-driven correction but omits the $p'^2E_x$ and $p'E_x^2$ terms. Relative flow-rate errors are ill-conditioned near stall, and the rigid-channel stall head is therefore compared instead. At $K=10$ and $h=E_x=1$, omission of only the mixed terms overestimates the stall head by $5.3 \%$, $11.1 \%$ and $27.7 \%$ for $g=0.125$, $0.28$ and $1$, respectively. Treating the pressure-driven branch as Newtonian increases these errors to $6.6 \%$, $14.8 \%$ and $51.1 \%$. At $g=1$ and $r=0.75$, the corresponding pressure-gradient errors are approximately $70$--$73 \%$. These differences quantify the error of the superposed approximations relative to the full reduced mixed-flow relation; they are not an independent validation of the constitutive model. The full relation is used throughout the loaded calculations.
	
	\subsection{Validation of the reduced transport model}
	\label{sec:transport_validation}
	
	The reduced transport model is tested against Brownian-dynamics simulations using the same lubrication velocity field, wall law and electrostatic and constitutive approximations, but without the Taylor--Aris or multiple-scales reduction. The particles retain their transverse positions and sample the local shear directly. Comparison of their exit-time moments with the backward-equation predictions therefore assesses the axial averaging and transverse-equilibration approximation for the specified flow and boundary conditions. Particle trajectories are advanced using the Euler--Maruyama scheme \citep{Higham2001Algorithmic},
	\begin{align}
		\mathrm{d}x
		&=u(x,y) \mathrm{d}t
		+\sqrt{\frac{2\delta}{\mathrm{Pe}}} \mathrm{d}W_x,
		\\
		\mathrm{d}y
		&=v(x,y) \mathrm{d}t
		+\sqrt{\frac{2}{\mathrm{Pe} \delta}} \mathrm{d}W_y,
		\label{eq:brownian_sde}
	\end{align}
	where $W_x$ and $W_y$ are independent Wiener processes. Continuity and centreline symmetry give
	\begin{equation}
		v(x,y)
		=-\int_0^y\partial_xu(x,y') \mathrm{d}y',
		\qquad
		u(x,y)=U \left(x,\eta\right),
		\qquad
		\eta=\frac{y}{h(x)}.
		\label{eq:transverse_velocity}
	\end{equation}
	
	Particles are uniformly released across the reflecting inlet and absorbed at $x=1$; the channel walls are also reflecting. In the scaled coordinates, the molecular-diffusion tensor is proportional to $\operatorname{diag}(\delta/\mathrm{Pe},1/(\mathrm{Pe}\delta))$. Thus, for the upper wall $F=y-h(x)=0$, no-flux reflection is conormal, with direction $\boldsymbol{D}\boldsymbol{\nabla}F\propto(-\delta^2h',1)$. Exit times yield particle estimates of $t_R$, $\sigma_t^2$ and $N=t_R^2/\sigma_t^2$. Relative discrepancies are measured against the particle estimates as $e_A=100|A_{\mathrm{red}}-A_{\mathrm{BD}}|/A_{\mathrm{BD}}$, for $A=t_R,N$. The six cases in table~\ref{tab:valid} test free flow and two loaded throughputs, Newtonian and viscoelastic flow, and selected EDL thicknesses and solute diffusivities. The model entries are evaluated from the full reflecting--absorbing backward problem, rather than its advection-dominated approximation. From the displayed values, the largest mean-transit-time discrepancy is approximately $0.32 \%$, while the plate-number discrepancy ranges from $0.34 \%$ to $2.17 \%$. The reduced model slightly underpredicts $t_R$ and overpredicts $N$ in all six cases. The largest plate-number difference occurs for the loaded Newtonian case and is approximately $2.2$ times the reported standard error of the Brownian-dynamics estimate. The standard errors quantify particle-sampling uncertainty; they are not error bounds for the reduced model.
	
	The additional comparisons in figure~\ref{fig:optimum}(b), at $K=10$, $r=0.5$, $g=0.125$, $C_m=0.1$ and $\delta=10^{-3}$ under CC operation, give plate-number discrepancies of $4.1 \%$, $1.7 \%$ and $0.3 \%$ for $\mathrm{Pe}=20,50,100$, respectively. The $2.2 \%$ agreement quoted for table~\ref{tab:valid} therefore applies to those six benchmarks, rather than to every comparison. These tests assess mean transit and band broadening through the first two exit-time moments; they do not establish agreement of the complete exit-time distribution.
	
	\begin{table}
		\centering
		\caption{Comparison of the reduced model with Brownian-dynamics simulations under CC operation, using $\epsilon_p=0.5$, $C_m=0.1$, $\delta=0.01$, $2\times10^4$ particles and time step $2\times10^{-3}$. Model values use the full backward equations \eqref{eq:backward}. Uncertainties in the particle estimates of $N$ denote one standard error; $e_N$ is the relative plate-number discrepancy in per cent, evaluated from the displayed values.}
		\label{tab:valid}
		\begin{tabular}{ccccccccc}
			\toprule
			$r$ & $K$ & $\mathrm{De}_{\kappa}$ & $\mathrm{Pe}$
			& $t_R$ particles & $t_R$ model
			& $N$ particles & $N$ model & $e_N$ \\
			\midrule
			$1.00$ & $10$ & $0.5$ & $20$ & $1.022$ & $1.019$ & $523.7\pm5.2$ & $525.5$ & $0.34$ \\
			$0.75$ & $10$ & $0.5$ & $20$ & $1.405$ & $1.402$ & $621.5\pm6.2$ & $629.2$ & $1.24$ \\
			$0.50$ & $10$ & $0.5$ & $20$ & $2.166$ & $2.160$ & $363.7\pm3.6$ & $366.2$ & $0.69$ \\
			$0.50$ & $5$  & $0.5$ & $20$ & $2.403$ & $2.396$ & $372.9\pm3.7$ & $379.1$ & $1.66$ \\
			$0.50$ & $10$ & $0.0$ & $20$ & $2.359$ & $2.353$ & $327.4\pm3.3$ & $334.5$ & $2.17$ \\
			$0.50$ & $10$ & $0.5$ & $10$ & $2.165$ & $2.158$ & $214.5\pm2.1$ & $216.7$ & $1.03$ \\
			\bottomrule
		\end{tabular}
	\end{table}
	
	Analytical profiles provide separate benchmarks for the cell problem. The plane-Poiseuille profile gives $\mathcal K=2\bar u^2h^2/105$ \citep{Datta2008Dispersion}, while the loaded thin-EDL parabolic core gives \eqref{eq:anchor}. These results check the coefficient normalisation and the response to a known shear profile. The finite-$K$ approach to the limiting core formula is a separate asymptotic comparison; its weak-loading crossover is described in Section~\ref{sec:thin_EDL_crossover}. A further flow benchmark is the Newtonian rigid-channel stall pressure,
	\begin{equation}
		p_{\mathrm{stall}}
		=3\left(1-\frac{\tanh K}{K}\right),
	\end{equation}
	which approaches $3$ as $K\to\infty$, corresponding to $\Delta p_{\mathrm{stall}}^*=3\eta_0U_{\mathrm{hs}}L/h_0^2$.
	
	\subsection{Numerical convergence}
	\label{sec:numerical_convergence}
	
	Convergence checks address the numerical evaluation of the reduced equations separately from the transport-reduction comparison above. The local flow-rate relation is inverted by Newton iteration using \eqref{eq:monotone}; the normalised residual is reduced below $10^{-13}$. The pressure equation is integrated by a fourth-order Runge--Kutta method on 121 axial nodes. Refinement to 481 nodes changes the integrated wall loading by $6\times10^{-6}$, providing a check on the axial pressure integration for this observable. Under CV operation, the conserved field constant is iterated until the absolute voltage residual $\left|\int_0^1E_x \mathrm{d}x-1\right|$ is below $10^{-12}$. The cell problem is evaluated using trapezoidal quadrature with 1601 transverse nodes and remains stable to six significant figures under refinement. The backward moment equations are solved on the converged axial fields. Flow-rate and current conservation are imposed by construction and therefore serve as consistency constraints rather than as independent validation data.
	
	For the six benchmark states in table~\ref{tab:valid}, the backward equations are evaluated in conservative slope form. With $D=D_{\mathrm{ax}}$, $z=AD T'$ and $w=AD V'$, they become $z'=-(\bar u/D)z-A$ and $w'=-(\bar u/D)w-2AD[T']^2$, with $z(0)=w(0)=0$. The outlet conditions then give $T(0)=-\int_0^1z/(AD) \mathrm dx$ and $V(0)=-\int_0^1w/(AD) \mathrm dx$. This formulation includes the diffusivity-gradient and area-drift terms without numerical differentiation. Fourth-order Runge--Kutta integration uses 4000 axial intervals for the pressure field and 40000 for the moments, with linear interpolation of the flow fields and 80-point Gauss--Legendre quadrature for the transverse integrals. Doubling both axial resolutions from 2000 and 20000 intervals changes $t_R$ by at most $2.2\times10^{-9}$ and $N$ by at most $4.1\times10^{-7}$ over these six states. For the uniform free-flow benchmark, the constant-drift, constant-diffusivity solution of \eqref{eq:backward} gives $t_R=1.01886$ and $N=525.54$, in agreement with table~\ref{tab:valid} at the reported precision.
	
	Particle velocities are interpolated bilinearly from a $61\times161$ grid in $(x,\eta)$. Halving the time step and varying the particle number from $4\times10^3$ to $2\times10^4$ change $N$ by less than $1 \%$ in the reported sensitivity checks. Omitting the transverse velocity changes $N$ by less than $1 \%$, while the comparison of conormal and transverse wall reflection changes it by at most $0.2 \%$. These tests quantify sensitivity to particle discretisation and boundary treatment for the cases considered; they do not extend the regime of validity of the underlying flow model.
	
	\subsection{Parameter values and regime checks}
	\label{sec:regime_checks}
	
	The dimensional reference set uses $h_0=1~\mu\mathrm{m}$, $L=1~\mathrm{mm}$, $h_c=2~\mu\mathrm{m}$, $E_0=100~\mathrm{kV m^{-1}}$, $\zeta_0=-25~\mathrm{mV}$, $\eta_0=10^{-3}~\mathrm{Pa s}$, $\lambda=29~\mu\mathrm{s}$ and $D_s^*=10^{-10}~\mathrm{m^2 s^{-1}}$. The complete properties and ranges are listed in the supplementary material. The dimensional ranges are linked by the physical relations below. The results include the sensitivity cases and bulk-Ohmic extrapolations identified in the regime checks. For a symmetric monovalent electrolyte with equal ionic diffusivities, $\sigma=\varepsilon_eD_i^*\kappa^2$. At fixed permittivity, ionic diffusivity and temperature, varying the dimensional Debye length changes both $\sigma$ and $c_\infty$ in proportion to $\kappa^2$. The quoted conductivity and concentration are reference values at $K=10$. The reference set therefore corresponds to $c_\infty\simeq9.1\times10^{-6}~\mathrm{M}$, $U_{\mathrm{hs}}=1.74\times10^{-3}~\mathrm{m s^{-1}}$ and $p_0=1.74~\mathrm{kPa}$. With $\beta=2$, $C_m=0.1$ corresponds to a constrained modulus of $34.8~\mathrm{kPa}$.
	
	The local wall law represents a compressible, drained coating bonded to a rigid support \citep{Skotheim2005Soft}. At the largest compliance considered, the maximum dimensionless deflection and constrained strain are approximately $0.21$ and $0.10$, respectively. The $C_m=0.1$ results are therefore treated as a sensitivity case near the limit of the small-deformation approximation. Nearly incompressible or undrained poroelastic coatings lie outside the present model. At $|\zeta_0|=25~\mathrm{mV}$ and $T_{\mathrm{abs}}=293~\mathrm{K}$, $|e\zeta_0|/(k_{\mathrm B}T_{\mathrm{abs}})=0.99$. The wall zeta potential is therefore comparable in magnitude to the thermal voltage, and the dimensional estimates are reference values within the linearised model. Surface conductance estimated from the Bikerman--Dukhin relation \citep{Bazant2004Diffuse} gives $K_s=5.0\times10^{-12}~\mathrm{S}$ per wall at $K=10$. Hence $\mathrm{Du}=K_s/(\sigma h_0)$ is $0.036$ at $K=10$, $0.060$ near the plate-number optimum at $K\simeq6$, and $0.18$ at $K=2$. Results below approximately $K=3$ are therefore identified as extrapolations of the bulk-Ohmic model. A local surface-conductance correction would replace $hE_x=\mathcal J$ with $(h+\mathrm{Du})E_x=\mathcal J$. Within the linearised slit model, the equilibrium electrical normal traction is bounded by
	\begin{equation}
		\Pi_{\mathrm{el}}^*
		=\frac{\varepsilon_e\kappa^2\zeta_0^2}
		{2\cosh^2(Kh)}.
	\end{equation}
	For the reference set, $\Pi_{\mathrm{el}}^*/p_0=1.1\times10^{-10}$ at $K=10$ and remains below $5.5\times10^{-5}$ over the reported range; its associated wall displacement is therefore negligible. The Reynolds number, normal-stress correction, wall slope, streaming current, Joule heating and electrical wall traction are all small. The largest wall deformation, transverse-equilibration parameter, Debye--H\"uckel parameter and Dukhin number at small $K$ are less well separated from unity; these cases are identified explicitly and interpreted as sensitivity limits of the reduced model.
	
	\begin{table}
		\centering
		\caption{Approximation checks for the reference set at $K=10$, $C_m=0.1$, $\beta=2$, $\epsilon_p=0.5$ and $\mathrm{De}_{\kappa}=0.5$ under CC operation. Unless stated otherwise, entries refer to the specified reference case, with flow-dependent values maximised over its forward-pumping branch; additional validation settings and values of $K$ are indicated in the table.}
		\label{tab:regime}
		\resizebox{\textwidth}{!}{
			\begin{tabular}{lll}
				\toprule
				Quantity & Definition & Value \\
				\midrule
				$\mathrm{Re}$
				& $\rho_fU_{\mathrm{hs}}h_0/\eta_0$
				& $1.7\times10^{-3}$ \\
				$\mathrm{Re}_{\max}$
				& $\mathrm{Re}\max(|u|h)$
				& $1.9\times10^{-3}$ \\
				$\delta$
				& $h_0/L$
				& $10^{-3}$ \\
				$\delta \mathrm{Wi}_{\mathrm{loc}}$
				& $\delta\lambda\max|\partial u^*/\partial y^*|$
				& $6.3\times10^{-4}$ \\
				$\chi$
				& $\max|h-1|$
				& $0.21$ \\
				$\varepsilon_{\mathrm{w},\max}$
				& $\max|h-1|/\beta$
				& $0.10$ \\
				$\delta\max|h'|$
				& maximum wall slope
				& $2.8\times10^{-4}$ \\
				$\mathrm{Pe} \delta$
				& reference equilibration parameter (validation cases included)
				& $0.017$--$0.20$ \\
				$\mu_{\max}$
				& $\mathrm{Pe} \delta\max(U_{\mathrm{loc}}h^2)$
				& $\simeq0.019$ (dimensional reference) \\
				$|e\zeta_0|/(k_{\mathrm B}T_{\mathrm{abs}})$
				& Debye--H\"uckel parameter
				& $0.99$ \\
				$R_{\mathrm{stream}}$
				& streaming-to-conduction-current estimate
				& $6.0\times10^{-3}$ \\
				$\Delta T_{\max}$
				& $\sigma E_0^2h_0^2/(2k_f)$
				& $1.2\times10^{-6}~\mathrm{K}$ \\
				$\mathrm{Du}$
				& $K_s/(\sigma h_0)$
				& $0.036$ at $K=10$; $0.18$ at $K=2$ \\
				$\mathcal T_{\mathrm{el}}$
				& $\Pi_{\mathrm{el}}^*/p_0$
				& $1.1\times10^{-10}$ at $K=10$ \\
				\bottomrule
			\end{tabular}%
		}
	\end{table}

	\section{Results and discussion}\label{sec:results}
	
	We evaluate pumping and solute transport on the same loaded flow state. Shading or hatching for $K<3$ marks extrapolations of the bulk-Ohmic model (Section~\ref{sec:regime_checks}). 
	
	\subsection{Pumping characteristics and wall loading}\label{sec:pump}
	
\begin{figure}
		\centering
		\includegraphics[width=0.3\textwidth]{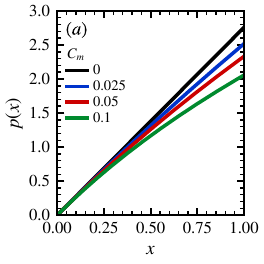}
		\includegraphics[width=0.3\textwidth]{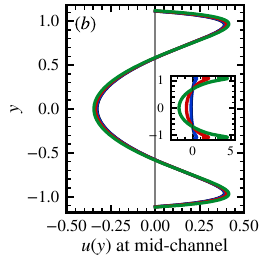}
		\includegraphics[width=0.3\textwidth]{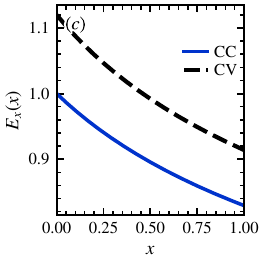}
		\caption{Stall states at $K=10$ and $\epsilon_p=0.5$. (a) CC pressure profiles for $\mathrm{De}_{\kappa}=0.5$ and $C_m=0,0.025,0.05,0.1$. (b) CC mid-channel velocity profiles for $C_m=0.1$ and $\mathrm{De}_{\kappa}=0,0.25,0.5,0.75$; inset: $10^2(u-u_{\mathrm N})$. (c) CC and CV electric fields for $\mathrm{De}_{\kappa}=0.5$ and $C_m=0.1$, on their respective stall geometries.}
		\label{fig:stall}
	\end{figure}

	The pressure-free curves in figure~\ref{fig:flow} set the throughput reference. Increasing $K$ reduces the EDL velocity deficit, while the thin-EDL rheological enhancement enters through the effective slip (Section~\ref{sec:reference_state} and \S S2 of the supplementary material) \citep{Afonso2009Analytical}. Unless stated otherwise, the loaded pump results use constant-current (CC) operation and $\epsilon_p=0.5$. At stall, electroosmotic forcing is balanced by an adverse pressure gradient. Compliance is the dominant influence in the reference case $K=10$: increasing $C_m$ from zero to $0.1$ lowers the stall head from $2.76$ to $2.06$, a reduction of $25 \%$, while the outlet displacement reaches $h_1=0.21$ (figure~\ref{fig:stall}(a)). Wall expansion increases the pressure-driven conductance, which scales approximately as $h^3$, while weakening the local CC field by a factor of $h^{-1}$.
	
	At stall, electroosmotic forcing drives fluid forward near the walls, while the adverse pressure gradient produces a return flow in the core (figure~\ref{fig:stall}(b)). This balance between electroosmotic forcing and an opposing pressure gradient underlies electroosmotic pumping \citep{Laser2004Review}. At $K=10$ and $C_m=0.1$, increasing $\mathrm{De}_{\kappa}$ from zero to $0.75$ raises the stall head by approximately $3 \%$. The change in stall head is therefore small in this reference case. The corresponding coefficient is $a=2g/K^2\leq5.6\times10^{-3}$, and the inset resolves the velocity difference from the Newtonian profile.
	
\begin{figure}
		\centering
		\includegraphics[width=0.49\textwidth]{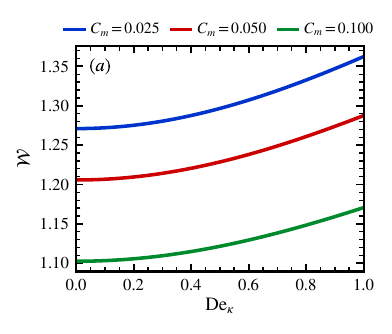}
		\hfill
		\includegraphics[width=0.49\textwidth]{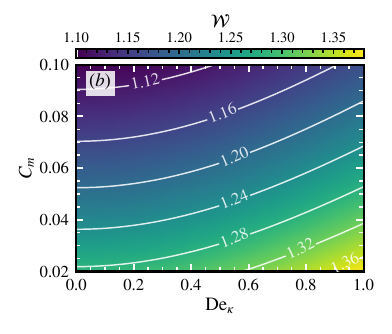}
		\caption{Integrated pressure load $\mathcal W$ at stall for $K=10$ and $\epsilon_p=0.5$ under CC operation. (a) $\mathcal W$ versus $\mathrm{De}_{\kappa}$ for $C_m=0.025,0.050,0.100$. (b) $\mathcal W$ over $(\mathrm{De}_{\kappa},C_m)$.}
		\label{fig:capacity}
	\end{figure}

	Figure~\ref{fig:stall}(c) compares the CC and CV fields on their respective self-consistent stall geometries. The electrical constraint modifies the pressure the pump sustains against the hydraulic load via the field normalisation in \eqref{eq:field_closures}. The integrated wall load $\mathcal W$, defined in Section~\ref{sec:operating}, corresponds to a dimensional pressure force $p_0L\mathcal W$ per unit span on either wall. At $K=10$, increasing $\mathrm{De}_{\kappa}$ from zero to unity raises $\mathcal W$ by approximately $6$--$7 \%$ over the compliance range in figure~\ref{fig:capacity}(a). This increase is smaller than predicted by a superposed-flow approximation because nonlinear mixed terms reduce the viscoelastic gain in the presence of an adverse pressure gradient. Increasing compliance lowers $\mathcal W$ through the field and conductance changes described for figure~\ref{fig:stall}.
	
	\begin{figure}
		\centering
		\includegraphics[width=0.6\textwidth]{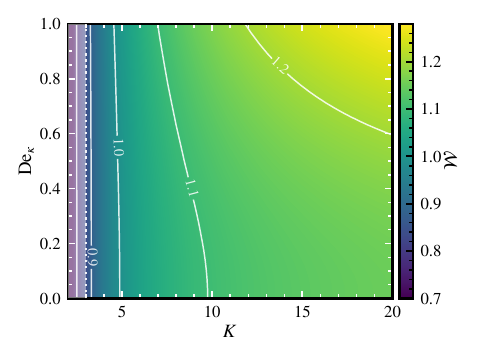}
		\caption{Integrated pressure load $\mathcal W$ at stall over $(K,\mathrm{De}_{\kappa})$ at $C_m=0.1$ and $\epsilon_p=0.5$ under CC operation. Shading: $K<3$.}
		\label{fig:loadmap}
	\end{figure}
    
\begin{figure}
		\centering
		\includegraphics[width=0.49\textwidth]{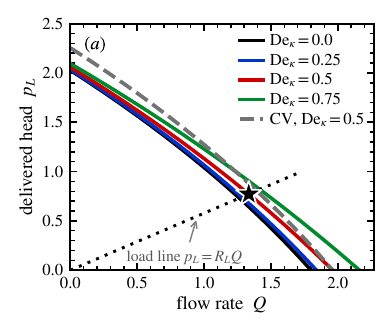}
		\hfill
		\includegraphics[width=0.49\textwidth]{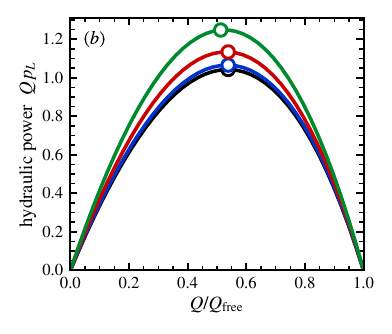}
		\caption{Pump characteristics at $K=10$, $C_m=0.1$ and $\epsilon_p=0.5$. (a) Head versus flow rate: CC curves for $\mathrm{De}_{\kappa}=0,0.25,0.5,0.75$ (solid), CV curve at $\mathrm{De}_{\kappa}=0.5$ (dashed), and load line (dotted). Star: CC operating point at $\mathrm{De}_{\kappa}=0.5$. (b) CC hydraulic power $Qp_L$ versus $r$, with the same colours as (a). Open circles: maxima.}
		\label{fig:pump}
	\end{figure}

	In the reference map of figure~\ref{fig:loadmap}, the integrated stall load rises with $K$ and $g$: the stronger electroosmotic drive supports a larger adverse pressure gradient. For the reference parameters in figure~\ref{fig:pump}(a), the delivered head decreases monotonically between stall and free flow, and the load-line intersection selects the operating point. Viscoelasticity shifts the characteristic upwards, increasing both the stall head and the pressure delivered at fixed throughput. The hydraulic power in figure~\ref{fig:pump}(b) vanishes at both endpoints and is greatest at intermediate throughput, where pressure delivery and flow rate remain finite. The throughput fraction at maximum power varies weakly over the rheological range considered.
	
	\subsection{Solute dispersion and plate-number performance}\label{sec:disp}
	
	Loading changes both the velocity differences that produce Taylor--Aris spreading and the mean speed that sets the transit time. We first examine the shear-dispersion coefficient, then the plate number, which together measure their combined effect; all figures in this subsection use the CC operation. The comparison of the CC and CV thin-EDL plateaux follows analytically from \eqref{eq:thin_EDL_compliant_plateau}. The pressure-free limits are given in \S S4 of the supplementary material; the loaded response is governed by the shear balance in Section~\ref{sec:axial_dispersion}. Opposing pressure-driven flow can reduce dispersion in rigid Newtonian tubes \citep{Samuel2025Taylor}, and mixed-flow viscoelastic dispersion can be non-monotonic in the forcing ratio \citep{Dejam2026Reduced}. Here the hydraulic load also couples the pressure gradient to wall deformation and the electric field.
	
\begin{figure}
		\centering
		\includegraphics[width=0.49\textwidth]{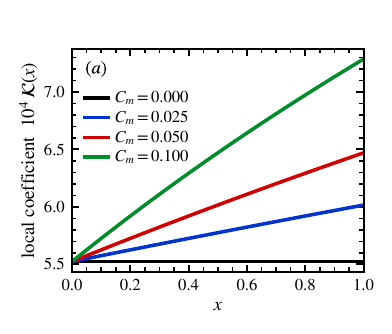}
		\hfill
		\includegraphics[width=0.49\textwidth]{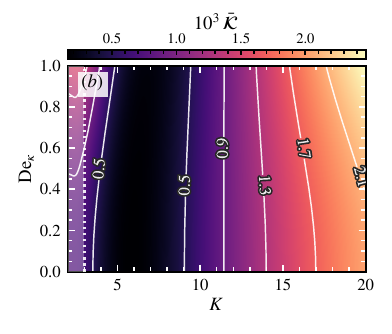}
		\caption{Loaded dispersion at $r=0.5$ and $\epsilon_p=0.5$ under CC operation. (a) $\mathcal K(x)$ at $K=10$ and $\mathrm{De}_{\kappa}=0.5$ for $C_m=0,0.025,0.05,0.1$. (b) $\overline{\mathcal K}$ over $(K,\mathrm{De}_{\kappa})$ at $C_m=0.1$. Shading: $K<3$.}
		\label{fig:disp}
	\end{figure}

	Compliance couples dispersion to the loaded profiles $h(x)$, $E_x(x)$ and $u(x,y)$. In the reference cases of figure~\ref{fig:disp}(a), $\mathcal K$ increases towards the expanding outlet at $K=10$. Expansion changes the transverse diffusion distance, the local electric field and the adverse pressure gradient together, thereby shifting the balance between electroosmotic and pressure-driven shear. The observed increase is a finite-$K$ result, not a general consequence of a larger gap. In the fixed-$g$ thin-EDL limit under CC operation, the Newtonian plateau is independent of compliance, while expansion lowers the local plateau for $g>0$ on the positive-pressure-gradient branch, as shown in Section~\ref{sec:thin_EDL_compliance}. CV operation changes this balance through the global field normalisation. Figure~\ref{fig:disp}(b) summarises the finite-$K$ response using the path average
    
	\begin{equation}
		\overline{\mathcal K}
		=\int_0^1\mathcal K(x) \mathrm{d}x
	\end{equation}
	over the $(K,\mathrm{De}_{\kappa})$ plane. This spatial average is a useful diagnostic; the transit-time weighting relevant to plate number is considered below.
	
\begin{figure}
		\centering
		\includegraphics[width=0.49\textwidth]{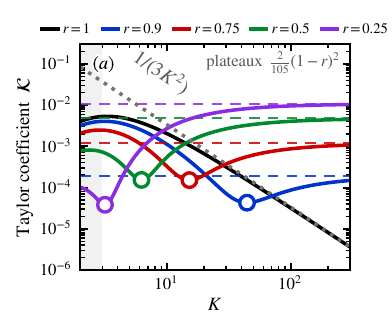}
		\hfill
		\includegraphics[width=0.49\textwidth]{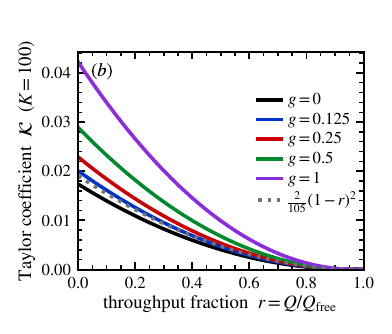}
		\caption{Taylor--Aris coefficient in a rigid slit under CC operation. (a) Newtonian coefficient versus $K$ for the throughput fractions shown. Dotted curve: free-flow asymptote; dashed curves: \eqref{eq:rigid_thin_EDL_dispersion}; open circles: minima. (b) Coefficient versus $r$ at $K=100$ for the indicated $g$. Shading: $K<3$.}
		\label{fig:loaddisp}
	\end{figure}

	Figure~\ref{fig:loaddisp} isolates the loaded thin-EDL mechanism. In the free-flow state, the velocity deficit becomes confined to the walls; at fixed $r<1$, the adverse pressure gradient sustains the core shear and the coefficient approaches \eqref{eq:rigid_thin_EDL_dispersion}. The $(1-r)^2$ dependence follows from a velocity deficit that is linear in the throughput reduction and a dispersion coefficient that is quadratic in that deficit. For the case of weak loading, the crossover scale \eqref{eq:thin_EDL_crossover_scale} separates the $K^{-2}$ free-flow regime from the core-shear plateau. Near the crossover, the two contributions to the cell response partially cancel, producing the negative cross term in \eqref{eq:thin_EDL_weak_load} and the interior dispersion minimum. The balance estimate $K_c$ and the minimum location $K_{\min}$ are distinct. As $r\to1$, both move to larger $K$ values while the plateau vanishes quadratically. For higher loading, the weak-load estimate is replaced by the full-cell calculation. For the Newtonian reference curves in figure~\ref{fig:loaddisp}(a), the computed minimum moves from $K\simeq44$ at $r=0.9$ to $K\simeq6.2$ at $r=0.5$, with minimum coefficients approximately $1/4.4$ and $1/31$ of their respective plateau values. Rheology changes both the electroosmotic velocity and the pressure gradient required to maintain a prescribed throughput fraction. Its finite-$K$ effect therefore depends on the shear balance. In a rigid channel at $r=0.5$, increasing $g$ from zero to unity reduces $\mathcal K$ by about $6 \%$ at $K=9$, but raises it from $4.0\times10^{-3}$ to $9.9\times10^{-3}$ at $K=100$. These are reference-case comparisons. The general fixed-$g$ thin-EDL result is that the rigid-channel plateau increases as $(1+2g/3)^2$ at fixed $r$: rheology survives through the effective slip amplitude even though the leading core profile is Newtonian. It is to be noted that this limiting trend does not imply a monotonic response for intermediate values of $K$.
	
\begin{figure}
		\centering
		\includegraphics[width=0.49\textwidth]{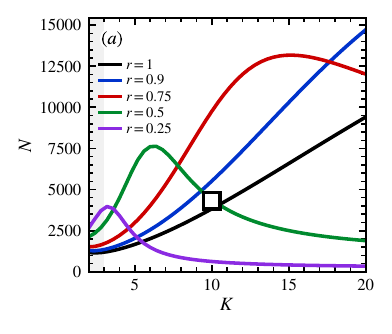}
		\hfill
		\includegraphics[width=0.49\textwidth]{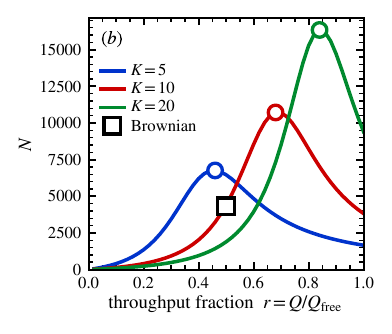}
		\caption{Plate number at $C_m=0.1$, $\mathrm{De}_{\kappa}=0.5$, $\epsilon_p=0.5$, $\mathrm{Pe}=50$ and $\delta=10^{-3}$ under CC operation. (a) $N$ versus $K$ for $r=1,0.9,0.75,0.5,0.25$. (b) $N$ versus $r$ for $K=5,10,20$, with open circles marking maxima. Open squares in both panels denote Brownian-dynamics estimates. Shading: $K<3$.}
		\label{fig:plates}
	\end{figure}

	The model curves for the plate number are obtained from the backward equations \eqref{eq:backward}. Open squares in figure~\ref{fig:plates} denote Brownian-dynamics estimates. The uniform, advection-dominated limit follows from Appendix~\ref{app:transit_moments}. In the reference cases of figure~\ref{fig:plates}(a), the free-flow plate number increases as the EDL becomes thinner, whereas each loaded curve has a maximum at finite $K$. Figure~\ref{fig:plates}(b) shows a corresponding optimum with respect to throughput: $r\simeq0.45$ at $K=5$ and $r\simeq0.84$ at $K=20$, with plate numbers approximately four and two times their free-flow values. Partial loading initially reduces the shear-induced spreading sufficiently to outweigh the loss of mean speed. In addition, the loading weakens directed transport and moves the velocity profile away from the strongest cancellation. The optimum plate number consequently depends on the molecular diffusion, mean speed and the axial distribution of dispersion. These optimum locations and gains are specific to the parameters, as depicted in the figure.
	
\begin{figure}
		\centering
		\includegraphics[width=0.49\textwidth]{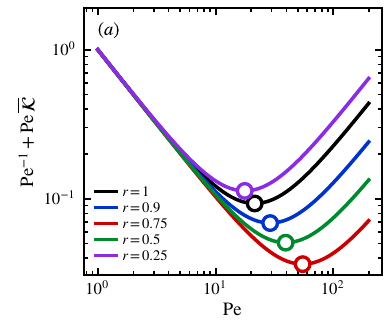}
		\hfill
		\includegraphics[width=0.49\textwidth]{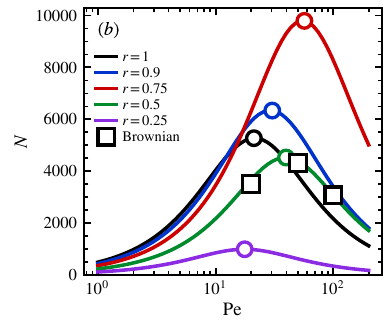}
		\caption{P\'eclet-number dependence at $K=10$, $C_m=0.1$, $\mathrm{De}_{\kappa}=0.5$, $\epsilon_p=0.5$ and $\delta=10^{-3}$ under CC operation, for $r=1,0.9,0.75,0.5,0.25$. (a) $\mathrm{Pe}^{-1}+\mathrm{Pe} \overline{\mathcal K}$; open circles: \eqref{eq:Pe_optimum}. (b) Plate number, with the same colours as (a); open circles: maxima; open squares: Brownian dynamics at $r=0.5$ and $\mathrm{Pe}=20,50,100$.}
		\label{fig:optimum}
	\end{figure}

	The P\'eclet number provides a second optimisation variable. At fixed hydrodynamics, the molecular contribution to dimensionless axial diffusion decreases with $\mathrm{Pe}$, while the Taylor--Aris contribution increases. An optimal P\'eclet number has also been obtained for rigid Newtonian tubes by minimising concentration variance over a prescribed transport distance \citep{Samuel2025Taylor}. The advection-dominated moment approximation in Appendix~\ref{app:transit_moments} gives
	\begin{equation}
		\mathrm{Pe}^{\star}
		=\mathcal K_t^{-1/2},
		\qquad
		\mathcal K_t=\frac{\int_0^1\mathcal K\bar u^{-3} \mathrm{d}x}
		{\int_0^1\bar u^{-3} \mathrm{d}x}.
		\label{eq:Pe_optimum}
	\end{equation}
	At this optimum, the molecular and shear contributions to the leading transit-time variance are equal. The factor $\bar u^{-3}$ gives greater weight to slower portions of the channel, explaining why minimising a spatially averaged dispersion coefficient alone does not determine separation performance. For the reference cases in figure~\ref{fig:optimum}, $\mathcal K_t$ differs from $\overline{\mathcal K}$ by less than $1 \%$, and the estimate from \eqref{eq:Pe_optimum} differs from the optimum obtained from the full backward equations by at most $0.4 \%$. These agreements quantify the accuracy of the advection-dominated estimate for those cases. The sweep varies solute diffusivity at fixed hydrodynamics and identifies the diffusivity that maximises the outlet plate number.
	
\begin{figure}
		\centering
		\includegraphics[width=0.6\textwidth]{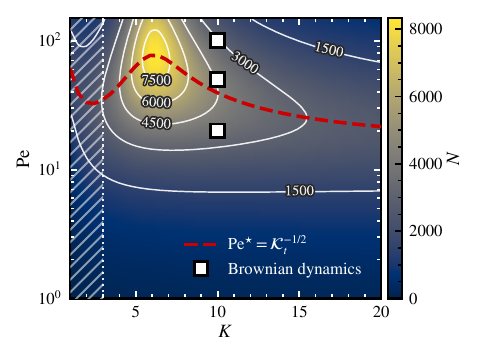}
		\caption{Plate number over $(K,\mathrm{Pe})$ at $r=0.5$, $C_m=0.1$, $\mathrm{De}_{\kappa}=0.5$, $\epsilon_p=0.5$ and $\delta=10^{-3}$ under CC operation. Dashed curve: \eqref{eq:Pe_optimum}; squares: Brownian dynamics; white contours: $N$; hatching: $K<3$.}
		\label{fig:sepmap}
	\end{figure}

	Figure~\ref{fig:sepmap} combines these balances in a reference map at $r=0.5$. Within the plotted parameter range, the plate number is maximum near $K\simeq6$ and $\mathrm{Pe}\simeq70$, where load-induced cancellation suppresses shear dispersion and molecular and shear spreading are comparable. The dashed curve follows from \eqref{eq:Pe_optimum}. This maximum is conditional on the specified throughput fraction, compliance and rheology. The dimensional controls are coupled: changing electrolyte concentration modifies $K$, conductivity and $\mathrm{De}_{\kappa}$, while changing the electric field modifies $\mathrm{Pe}$, $\mathrm{De}_{\kappa}$ and $C_m$. The map can therefore be interpreted as a design space traversed by jointly selecting the fluid, electrolyte, solute and electrical forcing.
	
	\subsection{Load--plate-number trade-off}\label{sec:tradeoff}
	
\begin{figure}
		\centering
		\includegraphics[width=0.49\textwidth]{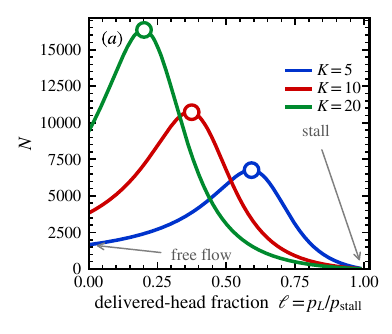}
		\hfill
		\includegraphics[width=0.49\textwidth]{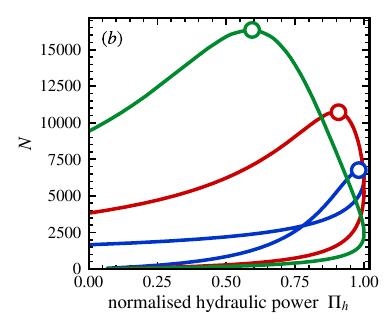}
		\caption{Load--plate-number trade-off at $C_m=0.1$, $g=0.125$, $\mathrm{Pe}=50$ and $\delta=10^{-3}$ under CC operation, for $K=5,10,20$. (a) $N$ versus $\ell=p_L/p_{\mathrm{stall}}$. (b) $N$ versus $\Pi_h=Qp_L/\max_r(Qp_L)$, normalised separately for each $K$, with the same colours as (a). Curves are parameterised by throughput; open circles mark maxima.}
		\label{fig:pareto}
	\end{figure}

	The pump and transport metrics depicted in figure~\ref{fig:pareto} correspond to the same operating state. Throughput parameterises each curve, with head fraction $\ell=p_L/p_{\mathrm{stall}}$ ranging from zero at free flow to unity at stall. Under the reflecting-inlet, absorbing-outlet formulation, the backward equations yield a finite stall value because molecular diffusion and shear-enhanced axial diffusion remain present at $K=10$, $N=1.56$. The maximum plate number occurs at finite head. The optimal head fractions are approximately $0.59$, $0.37$ and $0.20$ for $K=5$, $10$ and $20$, respectively. Figure~\ref{fig:pareto}(b) illustrates the same states in terms of normalised hydraulic power. When two states yield the same power, the state with the larger plate number dominates the other.
	
	Table~\ref{tab:opt} summarises the dependence of the optimum on $g$ and $C_m$. At $K=10$ and $C_m=0.1$, increasing $g$ from zero to unity raises $N_{\max}$ by approximately $38 \%$ and shifts $r_{\mathrm{opt}}$ from $0.70$ to $0.64$. At $K=10$, increasing $g$ from zero to unity raises $N_{\max}$ by at most approximately $40 \%$ across the tabulated compliance range. Increasing $C_m$ from $0.025$ to $0.1$ reduces $N_{\max}$ by approximately $1$--$2.5 \%$. At $C_m=0.1$, the optimal delivered-head fraction varies little with $g$ at either tabulated value of $K$.
	
	\begin{table}
		\centering
		\caption{Optimal throughput fraction $r_{\mathrm{opt}}$ and maximum plate number $N_{\max}$ obtained from \eqref{eq:backward} at fixed $K$, with $\mathrm{Pe}=50$ and $\delta=10^{-3}$. The gain is $N_{\max}/N_{\mathrm{free}}$ at $C_m=0.1$; $\ell_{\mathrm{opt}}$ is the corresponding delivered-head fraction.}
		\label{tab:opt}
		\begin{tabular}{lcccccccc}
			\toprule
			& \multicolumn{2}{c}{$C_m=0.025$}
			& \multicolumn{2}{c}{$C_m=0.050$}
			& \multicolumn{2}{c}{$C_m=0.100$}
			& & \\
			\cmidrule(lr){2-3}
			\cmidrule(lr){4-5}
			\cmidrule(lr){6-7}
			$g$
			& $r_{\mathrm{opt}}$ & $N_{\max}$
			& $r_{\mathrm{opt}}$ & $N_{\max}$
			& $r_{\mathrm{opt}}$ & $N_{\max}$
			& gain & $\ell_{\mathrm{opt}}$ \\
			\midrule
			\multicolumn{9}{l}{$K=10$} \\
			$0$     & $0.69$ & $10280$ & $0.69$ & $10237$ & $0.70$ & $10151$ & $2.74$ & $0.37$ \\
			$0.125$ & $0.68$ & $10897$ & $0.68$ & $10842$ & $0.69$ & $10737$ & $2.81$ & $0.37$ \\
			$0.5$   & $0.66$ & $12552$ & $0.66$ & $12464$ & $0.66$ & $12302$ & $3.01$ & $0.37$ \\
			$1$     & $0.64$ & $14365$ & $0.64$ & $14243$ & $0.64$ & $14018$ & $3.29$ & $0.38$ \\
			\midrule
			\multicolumn{9}{l}{$K=20$} \\
			$0$     & $0.84$ & $15403$ & $0.84$ & $15361$ & $0.85$ & $15278$ & $1.70$ & $0.20$ \\
			$0.125$ & $0.84$ & $16517$ & $0.84$ & $16459$ & $0.84$ & $16350$ & $1.74$ & $0.20$ \\
			$0.5$   & $0.84$ & $19613$ & $0.84$ & $19510$ & $0.84$ & $19313$ & $1.85$ & $0.20$ \\
			$1$     & $0.83$ & $23209$ & $0.83$ & $23050$ & $0.83$ & $22731$ & $1.99$ & $0.20$ \\
			\bottomrule
		\end{tabular}
	\end{table}
	
	The finite-$K$ optimum is conditional on throughput. Joint maximisation of $N$ over $r$ and $K$ moves the optimum towards the free-flow, thin-EDL boundary. For the reference case $g=0.125$, $\mathrm{Pe}=50$ and $\delta=10^{-3}$, the limiting plug-flow value is approximately $2.7\times10^4$. For these reference calculations, the ratio $N_{\max}/N_{\mathrm{free}}$ decreases from $4.08$ at $K=5$ to $1.74$ at $K=20$ and $1.05$ at $K=80$. The strongest benefit of partial loading therefore occurs for moderately thick EDLs.
	
	\subsection{Reduced-model structure and band shape}\label{sec:struct}

\begin{figure}
		\centering
		\includegraphics[width=0.49\textwidth]{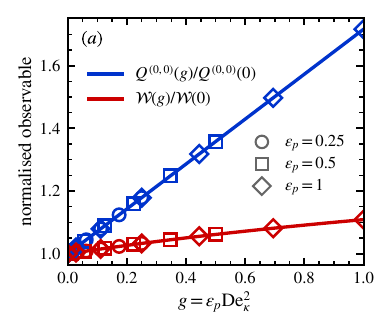}
		\hfill
		\includegraphics[width=0.49\textwidth]{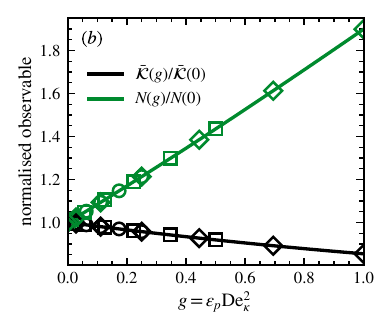}
		\caption{Observables versus $g=\epsilon_p\mathrm{De}_{\kappa}^{2}$ at $K=10$ and $C_m=0.1$ under CC operation, normalised by their Newtonian values. (a) Free-flow rate and stall pressure load. (b) $\overline{\mathcal K}$ and $N$ at $r=0.5$, $\mathrm{Pe}=50$ and $\delta=10^{-3}$. Symbols: $\epsilon_p=0.25,0.5,1$ with varying $\mathrm{De}_{\kappa}$; solid lines: model curves.}
		\label{fig:collapse}
	\end{figure}

    Within the retained leading-order equations, the flow and transport observables depend on $\epsilon_p$ and $\mathrm{De}_{\kappa}$ only through the combined parameter $g=\epsilon_p\mathrm{De}_{\kappa}^{2}$. Figure~\ref{fig:collapse} demonstrates this collapse for the free-flow rate, stall wall load, path-averaged dispersion coefficient and plate number. However, the separate validity condition involving $\delta \mathrm{Wi}_{\mathrm{loc}}$ must still be verified, as it retains an explicit dependence on the relaxation time.
	
	\begin{figure}
		\centering
		\includegraphics[width=0.49\textwidth]{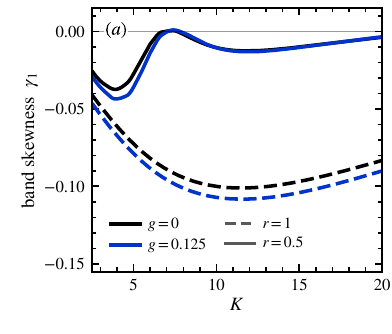}
		\hfill
		\includegraphics[width=0.49\textwidth]{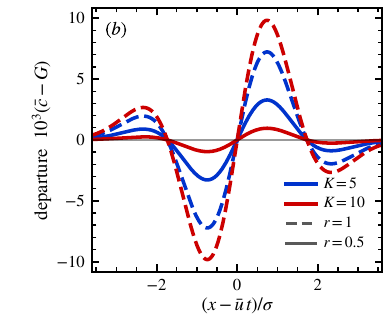}
		\caption{Spatial band shape from the auxiliary infinite-slit problem at $\mathrm{Pe}=50$ and $\delta=10^{-3}$ under CC operation, evaluated when the mean band position reaches $x=1$ ($t=1/\bar u$ in main-text units). (a) Leading spatial skewness $\gamma_1$ versus $K$ for $g=0,0.125$. (b) Leading Edgeworth reconstruction relative to the standard Gaussian $G$, at $g=0.125$ and $K=5,10$; $\bar c$ in this panel denotes the unit-mass density in the standardised spatial coordinate. Dashed curves: $r=1$; solid curves: $r=0.5$.}
		\label{fig:skew}
	\end{figure}
	
	The plate number $N$ uses the first two temporal exit-time moments and does not require a Gaussian arrival-time distribution. Figure~\ref{fig:skew} provides a complementary spatial description for an auxiliary, axially unbounded rigid slit. The moment hierarchy in \S S6 of the supplementary material gives the leading spatial skewness as $\gamma_1\sim C_SN_{\mathrm{sp}}^{-1/2}$, evaluated when the mean band position reaches the outlet location. Here $N_{\mathrm{sp}}=\bar u/(2D_{\mathrm{ax}})$ is the leading spatial sharpness measure defined in \S S6; it coincides with the advection-dominated approximation to $N$, rather than the exact reflecting--absorbing first-passage result. For the illustrated profiles, the spatial skewness is negative, indicating a right-tailed concentration. At $K=10$ and $g=0.125$, its magnitude decreases from approximately $0.107$ at free flow to $0.0105$ at half throughput. Loading therefore reduces the leading departure of the spatial band from a Gaussian in these cases. Viscoelasticity has a weaker effect at fixed throughput fraction. The next-order spatial kurtosis correction is $O(N_{\mathrm{sp}}^{-1})$, consistent with the long-time approach to normality analysed by \citet{Chatwin1970Normality}. These spatial statistics do not determine the skewness of the temporal outlet distribution. Quantitative energetic comparison of CV and CC operation would additionally require the electrical input power or electrohydrodynamic conversion efficiency and is not attempted here.
	
	\section{Conclusions}\label{sec:concl}
	
	We have examined electroosmotic pumping and passive-solute dispersion in a compliant slit containing a simplified Phan-Thien--Tanner fluid in the zero-Newtonian-solvent-viscosity limit. At fixed $g$ and a fixed throughput fraction $r=Q/Q_{\mathrm{free}}$ with $0<r<1$, the dimensionless shear-dispersion coefficient approaches a positive limit as the EDL thins. The adverse pressure gradient sustains core shear even when the flow there remains forward. At zero load, the coefficient decays as $K^{-2}$. The plateau is quadratic in the pressure-driven velocity deficit $U_s-\bar u$, as given by \eqref{eq:anchor}, and scales as $(1-r)^2$ in the rigid reference slit. In that slit, the weak-loading crossover shifts to larger $K$ as $r$ approaches unity, so the recovery of free-flow scaling is non-uniform. In the thin-EDL limit at fixed $g=\epsilon_p\mathrm{De}_{\kappa}^{2}$, the leading-order core velocity profile is parabolic, while rheology enters the plateau through the effective electroosmotic slip. In the rigid reference slit, the plateau scales as $(1+2g/3)^2$ at fixed $r$. At finite $K$, rheology also changes the adverse pressure gradient required to maintain the throughput fraction. Its effect on shear dispersion can therefore change sign with $K$.
	
	Compliance modifies the gap, pressure gradient and electric field together. At fixed $r$, the gap dependence of the Newtonian thin-EDL plateau cancels exactly under constant-current operation, although pressure and transit time change. Under constant-current operation, wall expansion lowers the local plateau on the positive-pressure-gradient branch relative to the rigid slit at the same $r$ and $g>0$. Under constant-voltage operation, expansion changes the global field normalisation and raises the Newtonian plateau above its rigid-channel value. The sign of the dispersion change thus depends on rheology and electrical operation; these limiting trends do not determine the response at finite $K$.
	
	At finite EDL thickness, partial loading can reduce shear dispersion sufficiently to increase the outlet plate number relative to free flow at the same thickness, despite the lower flow rate. The weak-loading crossover and the shear-dispersion minimum occur at distinct EDL thicknesses. The plate-number maximum also depends on molecular diffusion and transit-time weighting and can occur at a different thickness. For the loaded throughput fractions examined under constant current, the plate number peaks at finite $K$. When both $r$ and $K$ are varied in these calculations, the highest plate numbers in the range examined occur near the free-flow, thin-EDL boundary. In the reference calculations, moderately thick EDLs offer the strongest benefit of partial loading while delivering pressure. Design comparisons must therefore specify the required pressure or throughput together with rheology, compliance and electrical operation. 
	
	These conclusions apply within the lubrication, wall-strain, electrostatic and transverse-equilibration conditions in Section~\ref{sec:formulation}. The dimensional reference estimates use Debye--H\"uckel linearisation at a wall zeta potential comparable in magnitude to the thermal voltage. For the reference parameter set, results below approximately $K=3$ are extrapolations of the bulk-Ohmic model. The fixed-$g$ thin-EDL limit does not capture variations in electrolyte concentration alone. CV and CC comparisons use equal undeformed reference field, so no ranking at equal electrical power is implied. Brownian-dynamics comparisons using the same velocity field support the transport reduction, with plate-number agreement within $2.2 \%$ for six benchmarks. Experiments measuring loaded velocity profiles, pressure distributions and transit-time statistics could test the proposed mechanism. Extensions to finite-solvent viscosity, nonlinear electrostatics and non-local wall mechanics would further broaden the present framework.
	
	\backsection[Acknowledgements]{A. K. Nayak thanks the National Board for Higher Mathematics (NBHM), India, and the Science and Engineering Research Board (SERB), India (grant no.\ CRG/2023/006863), and the Anusandhan National Research Foundation (ANRF/ARGM/2025/002761/MTR), India, for their support during the preparation of this manuscript.}
	
	\backsection[Supplementary material]{Supplementary material comprising the derivations of the wall law, the reference-flow asymptotic limits, the mixed-flow relation, the Taylor--Aris coefficient, the consistency limits, the moment hierarchy, the thin-EDL asymptotics and the multiple-scales reduction, together with the dimensional reference properties and ranges.}
	
	\backsection[Funding]{A. K. Nayak received financial support from the Science and Engineering Research Board (SERB), India (grant no.\ CRG/2023/006863) and the Anusandhan National Research Foundation, India (ANRF/ARGM/2025/002761/MTR).}
	
	\backsection[Declaration of interests]{The authors report no conflict of interest.}
	
	\backsection[Data availability statement]{The data that support the findings of this study are available from the corresponding author upon reasonable request.}
	
	\backsection[Author ORCIDs]{Subhajyoti Sahoo, https://orcid.org/0000-0003-2538-8841; Ameeya Kumar Nayak, https://orcid.org/0000-0002-5040-1816; Kirti Chandra Sahu, https://orcid.org/0000-0002-7357-1141.}
	
	\backsection[Author contributions]{S. Sahoo: methodology, formal analysis, software, validation, writing -- original draft. A. K. Nayak: conceptualization, supervision, methodology, investigation, writing -- review and editing. K. C. Sahu: supervision, investigation, methodology, writing -- review and editing.}
	
	\appendix
	\titleformat{\section}{\normalfont\normalsize}{Appendix~\thesection.}{0.5em}{\bfseries}
	
	\section{Field expansion and pressure-free reference flow}
	\label{app:flow_details}
	
	\subsection{Small-deformation field expansion}
	\label{app:field_expansions}
	
	For small deformation, write $h(x)=1+\chi H_1(x)$, where $\chi\ll1$, $\max_x|H_1|=1$ and $\chi H_1=C_mp$. Define the axial average by $\langle f\rangle_x=\int_0^1f(x) \mathrm{d}x$. The electrical closures give
	\begin{align}
		E_x^{\mathrm{CC}}
		&=1-\chi H_1+O(\chi^2), \\
		E_x^{\mathrm{CV}}
		&=1+\chi\bigl(\langle H_1\rangle_x-H_1\bigr)+O(\chi^2).
		\label{eq:field_expansions}
	\end{align}
	Thus, the mean deformation contribution is removed by the global voltage normalisation under CV operation. These expansions are used for interpretation only; the loaded calculations use the full closures in \eqref{eq:field_closures}. The pure electroosmotic viscoelastic contribution is proportional to $E_x^3$. Hence, if $E_x=1+\chi e_1+O(\chi^2)$, then $E_x^3=1+3\chi e_1+O(\chi^2)$; the factor of three must therefore be retained in the first-order deformation correction.
	
	\subsection{Undeformed pressure-free solution}
	\label{app:reference_state}
	
	For the undeformed, pressure-free reference state, $h=1$, $p'=0$ and $E_x=1$. Within the Debye--H\"uckel and lubrication approximations \citep{Afonso2009Analytical}, the sPTT relation with no slip, $u^{(0,0)}(\pm1)=0$, gives
	\begin{equation}
		u^{(0,0)}(y)
		=1-\frac{\cosh(Ky)}{\cosh K}
		+g 
		\frac{
			\cosh(3K)-\cosh(3Ky)-9\cosh K+9\cosh(Ky)
		}{
			6\cosh^3K
		}.
		\label{eq:u00}
	\end{equation}
	
	Integration across the gap gives \eqref{eq:Q00}, with
	\begin{equation}
		\mathcal{Q}_1(K)=\frac{4}{9K}
		\left[
		3K-\frac{9K}{\cosh^2K}
		+\frac{6\sinh K}{\cosh^3K}
		-\tanh^3K
		\right].
		\label{eq:Q1}
	\end{equation}
	The part of \eqref{eq:u00} independent of $g$ is the Newtonian electroosmotic profile, while the term proportional to $g$ is the sPTT correction. The integration steps and fixed-$g$ asymptotic limits are given in the supplementary material.
	
	\section{Dispersion and transit-time relations}
	\label{app:transport_details}
	
	\subsection{Rheological dependence at fixed local fields}
	\label{app:local_rheology}
	
	At fixed $K$ and fixed local $h$, $E_x$ and $p'$, the shear stress is independent of $g$, while the velocity and cell function are affine in $g$. With $\varphi=\varphi_0+g\varphi_1$, the dispersion coefficient is exactly
	\begin{equation}
		\mathcal{K}
		=\mathcal{K}_0
		+2g
		\left\langle
		\partial_y\varphi_0 
		\partial_y\varphi_1
		\right\rangle_y
		+g^2
		\left\langle
		\left(\partial_y\varphi_1\right)^2
		\right\rangle_y,
		\qquad
		\mathcal{K}_0
		=\left\langle
		\left(\partial_y\varphi_0\right)^2
		\right\rangle_y.
		\label{eq:Kquad}
	\end{equation}
	Along a prescribed-throughput characteristic, however, $p'$, $p$, $h$ and $E_x$ generally depend on $g$. The loaded coefficient is therefore not, in general, a quadratic polynomial in $g$ and is evaluated from the fully coupled solution. Within the retained leading-order velocity, flux and cell equations, $\epsilon_p$ and $\mathrm{De}_{\kappa}$ enter through $g$, while the validity condition involving $\mathrm{Wi}_{\mathrm{loc}}$ retains separate dependence on the relaxation time.
	
	\subsection{Advection-dominated moments and P\'eclet-number optimum}
	\label{app:transit_moments}
	
	For advection-dominated forward flow, with $\bar u>0$ and $\max_x(D_{\mathrm{ax}}/\bar u)\ll1$, the leading transit-time moments for the reflecting inlet and absorbing outlet are
	\begin{equation}
		t_R\simeq
		\int_0^1\frac{\mathrm{d}x}{\bar u(x)},
		\qquad
		\sigma_t^2\simeq
		\int_0^1
		\frac{
			2\delta\left(
			\mathrm{Pe}^{-1}
			+\mathrm{Pe} \mathcal{K}(x)
			\right)
		}{
			\bar u^3(x)
		} \mathrm{d}x.
		\label{eq:moments}
	\end{equation}
	These expressions reduce to the uniform-channel results when $h$ is constant and are not used near stall, where $\bar u\to0$. Their role in the plate-number estimates and P\'eclet-number optimum is discussed in Section~\ref{sec:disp}. At fixed hydrodynamics and for $\mathcal K_t>0$, minimising the leading variance gives the P\'eclet-number optimum \eqref{eq:Pe_optimum}. The integration and optimisation steps are given in the supplementary material. The plotted model values of the plate number are calculated from the full backward equations.
	
	\bibliographystyle{jfm}
	\bibliography{jfm}

\begin{thebibliography}{49}
\expandafter\ifx\csname natexlab\endcsname\relax\def\natexlab#1{#1}\fi
\def\au#1{#1} \def\ed#1{#1} \def\yr#1{#1}\def\at#1{#1}\def\jt#1{\textit{#1}}
  \def\bt#1{#1}\def\bvol#1{\textbf{#1}} \def\vol#1{#1} \def\pg#1{#1}
  \def\publ#1{#1}\def\arxiv#1{#1}\def\org#1{#1}\def\st#1{\textit{#1}}

\bibitem[Afonso {\em et~al.\/}(2009)Afonso, Alves \&
  Pinho]{Afonso2009Analytical}
{\sc \au{Afonso, A.~M.}, \au{Alves, M.~A.} \& \au{Pinho, F.~T.}} \yr{2009}
  \at{Analytical solution of mixed electro-osmotic/pressure driven flows of
  viscoelastic fluids in microchannels}.  \jt{Journal of Non-Newtonian Fluid
  Mechanics}  \bvol{159}~(1-3),  \pg{50--63}.

\bibitem[Arcos {\em et~al.\/}(2018)Arcos, M\'{e}ndez, Bautista \&
  Bautista]{Arcos2018Dispersion}
{\sc \au{Arcos, J.~C.}, \au{M\'{e}ndez, F.}, \au{Bautista, E.~G.} \&
  \au{Bautista, O.}} \yr{2018}  \at{Dispersion coefficient in an
  electro-osmotic flow of a viscoelastic fluid through a microchannel with a
  slowly varying wall zeta potential}.  \jt{Journal of Fluid Mechanics}
  \bvol{839},  \pg{348--386}.

\bibitem[Aris(1956)]{Aris1956}
{\sc \au{Aris, R.}} \yr{1956}  \at{On the dispersion of a solute in a fluid
  flowing through a tube}.  \jt{Proceedings of the Royal Society of London A}
  \bvol{235},  \pg{67--77}.

\bibitem[Bazant {\em et~al.\/}(2004)Bazant, Thornton \&
  Ajdari]{Bazant2004Diffuse}
{\sc \au{Bazant, M.~Z.}, \au{Thornton, K.} \& \au{Ajdari, A.}} \yr{2004}
  \at{Diffuse-charge dynamics in electrochemical systems}.  \jt{Physical Review
  E}  \bvol{70},  \pg{021506}.

\bibitem[Boyko \& Christov(2023)]{Boyko2023NonNewtonian}
{\sc \au{Boyko, E.} \& \au{Christov, I.~C.}} \yr{2023}  \at{Non-{N}ewtonian
  fluid-structure interaction: Flow of a viscoelastic {O}ldroyd-{B} fluid in a
  deformable channel}.  \jt{Journal of Non-Newtonian Fluid Mechanics}
  \bvol{313},  \pg{104990}.

\bibitem[Boyko {\em et~al.\/}(2019)Boyko, Eshel, Gommed, Gat \&
  Bercovici]{Boyko2019Elastohydrodynamics}
{\sc \au{Boyko, E.}, \au{Eshel, R.}, \au{Gommed, K.}, \au{Gat, A.~D.} \&
  \au{Bercovici, M.}} \yr{2019}  \at{Elastohydrodynamics of a pre-stretched
  finite elastic sheet lubricated by a thin viscous film with application to
  microfluidic soft actuators}.  \jt{Journal of Fluid Mechanics}  \bvol{862},
  \pg{732--752}.

\bibitem[Boyko \& Stone(2022)]{Boyko2022Pressure}
{\sc \au{Boyko, E.} \& \au{Stone, H.~A.}} \yr{2022}  \at{Pressure-driven flow
  of the viscoelastic {Oldroyd-B} fluid in narrow non-uniform geometries:
  analytical results and comparison with simulations}.  \jt{Journal of Fluid
  Mechanics}  \bvol{936},  \pg{A23}.

\bibitem[Boyko \& Stone(2026)]{Boyko2026Taylor}
{\sc \au{Boyko, E.} \& \au{Stone, H.~A.}} \yr{2026}  \at{Taylor--{A}ris
  dispersion in shear-rate-dependent fluid flows}.  \jt{Journal of Fluid
  Mechanics}  \bvol{1037},  \pg{R4}.

\bibitem[Chakraborty {\em et~al.\/}(2012)Chakraborty, Prakash, Friend \&
  Yeo]{Chakraborty2012Fluid}
{\sc \au{Chakraborty, D.}, \au{Prakash, J.~R.}, \au{Friend, J.} \& \au{Yeo,
  L.}} \yr{2012}  \at{Fluid-structure interaction in deformable microchannels}.
   \jt{Physics of Fluids}  \bvol{24}~(10).

\bibitem[Chandler \& Vella(2020)]{Chandler2020Validity}
{\sc \au{Chandler, T. G.~J.} \& \au{Vella, D.}} \yr{2020}  \at{Validity of
  {Winkler}'s mattress model for thin elastomeric layers: beyond {Poisson}'s
  ratio}.  \jt{Proceedings of the Royal Society A}  \bvol{476}~(2242),
  \pg{20200551}.

\bibitem[Chang \& Santiago(2023)]{Chang2023Taylor}
{\sc \au{Chang, R.} \& \au{Santiago, J.~G.}} \yr{2023}  \at{Taylor dispersion
  in arbitrarily shaped axisymmetric channels}.  \jt{Journal of Fluid
  Mechanics}  \bvol{976},  \pg{A30}.

\bibitem[Chatterjee {\em et~al.\/}(2022)Chatterjee, Nayak \&
  Weigand]{Chatterjee2022Effect}
{\sc \au{Chatterjee, A.}, \au{Nayak, A.~K.} \& \au{Weigand, B.}} \yr{2022}
  \at{Effect of electromigration dispersion and non-{N}ewtonian rheology of a
  charged solute in a microcapillary}.  \jt{Physics of Fluids}  \bvol{34}~(11).

\bibitem[Chatwin(1970)]{Chatwin1970Normality}
{\sc \au{Chatwin, P.~C.}} \yr{1970}  \at{The approach to normality of the
  concentration distribution of a solute in a solvent flowing along a straight
  pipe}.  \jt{Journal of Fluid Mechanics}  \bvol{43}~(2),  \pg{321--352}.

\bibitem[Cheung {\em et~al.\/}(2012)Cheung, Toda-Peters \& Shen]{Cheung2012In}
{\sc \au{Cheung, P.}, \au{Toda-Peters, K.} \& \au{Shen, A.~Q.}} \yr{2012}
  \at{In situ pressure measurement within deformable rectangular
  polydimethylsiloxane microfluidic devices}.  \jt{Biomicrofluidics}
  \bvol{6}~(2).

\bibitem[Christov {\em et~al.\/}(2018)Christov, Cognet, Shidhore \&
  Stone]{Christov2018Flow}
{\sc \au{Christov, I.~C.}, \au{Cognet, V.}, \au{Shidhore, T.~C.} \& \au{Stone,
  H.~A.}} \yr{2018}  \at{Flow rate--pressure drop relation for deformable
  shallow microfluidic channels}.  \jt{Journal of Fluid Mechanics}  \bvol{841},
   \pg{267--286}.

\bibitem[Datta \& Ghosal(2008)]{Datta2008Dispersion}
{\sc \au{Datta, S.} \& \au{Ghosal, S.}} \yr{2008}  \at{Dispersion due to wall
  interactions in microfluidic separation systems}.  \jt{Physics of Fluids}
  \bvol{20}~(1),  \pg{012103}.

\bibitem[Dehe {\em et~al.\/}(2021)Dehe, Rehm \& Hardt]{Dehe2021Hydrodynamic}
{\sc \au{Dehe, S.}, \au{Rehm, I.-S.} \& \au{Hardt, S.}} \yr{2021}
  \at{Hydrodynamic dispersion in {Hele-Shaw} flows with inhomogeneous wall
  boundary conditions}.  \jt{Journal of Fluid Mechanics}  \bvol{925},
  \pg{A11}.

\bibitem[Dejam \& Hassanzadeh(2026)]{Dejam2026Reduced}
{\sc \au{Dejam, M.} \& \au{Hassanzadeh, H.}} \yr{2026}  \at{Reduced-order model
  for solute transport in mixed electro-osmotic and pressure-driven flows of
  viscoelastic fluids in microchannels}.  \jt{Physical Review E}  \bvol{114},
  \pg{015105}.

\bibitem[Ding(2023)]{Ding2023Shear}
{\sc \au{Ding, L.}} \yr{2023}  \at{Shear dispersion of multispecies electrolyte
  solutions in the channel domain}.  \jt{Journal of Fluid Mechanics}
  \bvol{970},  \pg{A27}.

\bibitem[Ding(2025)]{Ding2025Longtime}
{\sc \au{Ding, L.}} \yr{2025}  \at{Long-time asymptotics of passive scalar
  transport in periodically modulated channels}.  \jt{Journal of Fluid
  Mechanics}  \bvol{1023},  \pg{A19}.

\bibitem[Dutta \& Beskok(2001)]{Dutta2001Combined}
{\sc \au{Dutta, P.} \& \au{Beskok, A.}} \yr{2001}  \at{Analytical solution of
  combined electroosmotic/pressure driven flows in two-dimensional straight
  channels: finite {Debye} layer effects}.  \jt{Analytical Chemistry}
  \bvol{73}~(9),  \pg{1979--1986}.

\bibitem[Gardiner(2009)]{Gardiner2009Stochastic}
{\sc \au{Gardiner, C.}} \yr{2009} {\em Stochastic Methods: A Handbook for the
  Natural and Social Sciences\/}, 4th edn.  \publ{Berlin, Heidelberg:
  Springer}.

\bibitem[Ghosal(2002)]{Ghosal2002Lubrication}
{\sc \au{Ghosal, S.}} \yr{2002}  \at{Lubrication theory for electro-osmotic
  flow in a microfluidic channel of slowly varying cross-section and wall
  charge}.  \jt{Journal of Fluid Mechanics}  \bvol{459},  \pg{103--128}.

\bibitem[Ghosal(2006)]{Ghosal2006Electrokinetic}
{\sc \au{Ghosal, S.}} \yr{2006}  \at{Electrokinetic flow and dispersion in
  capillary electrophoresis}.  \jt{Annual Review of Fluid Mechanics}
  \bvol{38},  \pg{309--338}.

\bibitem[Ghosal \& Chen(2012)]{Ghosal2012Electromigration}
{\sc \au{Ghosal, S.} \& \au{Chen, Z.}} \yr{2012}  \at{Electromigration
  dispersion in a capillary in the presence of electro-osmotic flow}.
  \jt{Journal of Fluid Mechanics}  \bvol{697},  \pg{436--454}.

\bibitem[Higham(2001)]{Higham2001Algorithmic}
{\sc \au{Higham, D.~J.}} \yr{2001}  \at{An algorithmic introduction to
  numerical simulation of stochastic differential equations}.  \jt{SIAM Review}
   \bvol{43}~(3),  \pg{525--546}.

\bibitem[Hoshyargar {\em et~al.\/}(2018)Hoshyargar, Talebi, Ashrafizadeh \&
  Sadeghi]{Sadeghi2018Hydrodynamic}
{\sc \au{Hoshyargar, V.}, \au{Talebi, M.}, \au{Ashrafizadeh, S.~N.} \&
  \au{Sadeghi, A.}} \yr{2018}  \at{Hydrodynamic dispersion by electroosmotic
  flow of viscoelastic fluids within a slit microchannel}.  \jt{Microfluidics
  and Nanofluidics}  \bvol{22},  \pg{4}.

\bibitem[Jha {\em et~al.\/}(2026)Jha, Gunny, McGraw, Amarouchene \&
  Salez]{Jha2026Taylor}
{\sc \au{Jha, A.}, \au{Gunny, M.}, \au{McGraw, J.~D.}, \au{Amarouchene, Y.} \&
  \au{Salez, T.}} \yr{2026} Taylor dispersion in a soft tube. \textit{arXiv
  preprint} arXiv:2604.05592.

\bibitem[Kang {\em et~al.\/}(2014)Kang, Roh \& Overfelt]{Kang2014Pressure}
{\sc \au{Kang, C.}, \au{Roh, C.} \& \au{Overfelt, R.~A.}} \yr{2014}
  \at{Pressure-driven deformation with soft polydimethylsiloxane ({PDMS}) by a
  regular syringe pump: challenge to the classical fluid dynamics by comparison
  of experimental and theoretical results}.  \jt{RSC Advances}  \bvol{4},
  \pg{3102--3112}.

\bibitem[Laser \& Santiago(2004)]{Laser2004Review}
{\sc \au{Laser, D.~J.} \& \au{Santiago, J.~G.}} \yr{2004}  \at{A review of
  micropumps}.  \jt{Journal of Micromechanics and Microengineering}  \bvol{14},
   \pg{R35--R64}.

\bibitem[Mahapatra {\em et~al.\/}(2025)Mahapatra, Ruangkriengsin, Stone \&
  Boyko]{Mahapatra2025Finite}
{\sc \au{Mahapatra, B.}, \au{Ruangkriengsin, T.}, \au{Stone, H.~A.} \&
  \au{Boyko, E.}} \yr{2025}  \at{Viscoelastic fluid flow in a slowly varying
  planar contraction: the role of finite extensibility on the pressure drop}.
  \jt{Journal of Fluid Mechanics}  \bvol{1009},  \pg{A12}.

\bibitem[Masliyah \& Bhattacharjee(2006)]{Masliyah2006electrokinetic}
{\sc \au{Masliyah, J.~H.} \& \au{Bhattacharjee, S.}} \yr{2006} {\em
  Electrokinetic and Colloid Transport Phenomena\/}.  \publ{Hoboken, NJ: John
  Wiley \& Sons}.

\bibitem[McDonald \& Whitesides(2002)]{McDonald2002Poly}
{\sc \au{McDonald, J.~C.} \& \au{Whitesides, G.~M.}} \yr{2002}
  \at{Poly(dimethylsiloxane) as a material for fabricating microfluidic
  devices}.  \jt{Accounts of Chemical Research}  \bvol{35},  \pg{491--499}.

\bibitem[Mukherjee {\em et~al.\/}(2022)Mukherjee, Dhar, DasGupta \&
  Chakraborty]{Mukherjee2022Electrokinetically}
{\sc \au{Mukherjee, S.}, \au{Dhar, J.}, \au{DasGupta, S.} \& \au{Chakraborty,
  S.}} \yr{2022}  \at{Electrokinetically augmented load bearing capacity of a
  deformable microfluidic channel}.  \jt{Physics of Fluids}  \bvol{34},
  \pg{082019}.

\bibitem[Phan-Thien \& Tanner(1977)]{PhanThien1977New}
{\sc \au{Phan-Thien, N.} \& \au{Tanner, R.~I.}} \yr{1977}  \at{A new
  constitutive equation derived from network theory}.  \jt{Journal of
  Non-Newtonian Fluid Mechanics}  \bvol{2},  \pg{353--365}.

\bibitem[Raj {\em et~al.\/}(2017)Raj, DasGupta \&
  Chakraborty]{Raj2017Hydrodynamics}
{\sc \au{Raj, M.~K.}, \au{DasGupta, S.} \& \au{Chakraborty, S.}} \yr{2017}
  \at{Hydrodynamics in deformable microchannels}.  \jt{Microfluidics and
  Nanofluidics}  \bvol{21},  \pg{1--12}.

\bibitem[Ramos-Arzola \& Bautista(2021)]{RamosArzola2021Fluid}
{\sc \au{Ramos-Arzola, L.} \& \au{Bautista, O.}} \yr{2021}  \at{Fluid
  structure-interaction in a deformable microchannel conveying a viscoelastic
  fluid}.  \jt{Journal of Non-Newtonian Fluid Mechanics}  \bvol{296},
  \pg{104634}.

\bibitem[Roy {\em et~al.\/}(2025)Roy, B\'{e}g \& Rana]{Roy2025Dispersion}
{\sc \au{Roy, A.~K.}, \au{B\'{e}g, O.~A.} \& \au{Rana, A.~S.}} \yr{2025}
  \at{Dispersion of neutral solutes in viscoelastic microflows under combined
  electroosmotic and pressure forcing}.  \jt{Physics of Fluids}  \bvol{37},
  \pg{082054}.

\bibitem[Rubin {\em et~al.\/}(2017)Rubin, Tulchinsky, Gat \&
  Bercovici]{Rubin2017Elastic}
{\sc \au{Rubin, S.}, \au{Tulchinsky, A.}, \au{Gat, A.~D.} \& \au{Bercovici,
  M.}} \yr{2017}  \at{Elastic deformations driven by nonuniform lubrication
  flows}.  \jt{Journal of Fluid Mechanics}  \bvol{812},  \pg{841--865}.

\bibitem[Sackmann {\em et~al.\/}(2014)Sackmann, Fulton \&
  Beebe]{Sackmann2014The}
{\sc \au{Sackmann, E.~K.}, \au{Fulton, A.~L.} \& \au{Beebe, D.~J.}} \yr{2014}
  \at{The present and future role of microfluidics in biomedical research}.
  \jt{Nature}  \bvol{507},  \pg{181--189}.

\bibitem[Sahoo \& Nayak(2026)]{Sahoo2026Electroosmotic}
{\sc \au{Sahoo, S.} \& \au{Nayak, A.~K.}} \yr{2026}  \at{Electroosmotic
  lubrication in constricted microchannels with a compliant wall and {DLVO}
  disjoining pressure}.  \jt{Physical Review Fluids}  \bvol{11},  \pg{064201}.

\bibitem[Samuel {\em et~al.\/}(2026)Samuel, Chang, Ma, Alessio \&
  Santiago]{Samuel2026Varying}
{\sc \au{Samuel, C.~J.}, \au{Chang, R.}, \au{Ma, K.}, \au{Alessio, B.~M.} \&
  \au{Santiago, J.~G.}} \yr{2026}  \at{Taylor dispersion in long straight
  channels of arbitrary and gradually changing size and shape}.  \jt{Journal of
  Fluid Mechanics}  \bvol{1040},  \pg{A52}.

\bibitem[Samuel {\em et~al.\/}(2025)Samuel, Chang, Ma \&
  Santiago]{Samuel2025Taylor}
{\sc \au{Samuel, C.~J.}, \au{Chang, R.}, \au{Ma, K.} \& \au{Santiago, J.~G.}}
  \yr{2025}  \at{Taylor dispersion for coupled electroosmotic and
  pressure-driven flows in all time regimes}.  \jt{Journal of Fluid Mechanics}
  \bvol{1011},  \pg{A36}.

\bibitem[Shidhore \& Christov(2018)]{Shidhore2018Static}
{\sc \au{Shidhore, T.~C.} \& \au{Christov, I.~C.}} \yr{2018}  \at{Static
  response of deformable microchannels: a comparative modeling study}.
  \jt{Journal of Physics: Condensed Matter}  \bvol{30},  \pg{054002}.

\bibitem[Skotheim \& Mahadevan(2005)]{Skotheim2005Soft}
{\sc \au{Skotheim, J.~M.} \& \au{Mahadevan, L.}} \yr{2005}  \at{Soft
  lubrication: The elastohydrodynamics of nonconforming and conforming
  contacts}.  \jt{Physics of Fluids}  \bvol{17}~(9),  \pg{092101}.

\bibitem[Sollier {\em et~al.\/}(2011)Sollier, Murray, Maoddi \&
  Di~Carlo]{Sollier2011Rapid}
{\sc \au{Sollier, E.}, \au{Murray, C.}, \au{Maoddi, P.} \& \au{Di~Carlo, D.}}
  \yr{2011}  \at{Rapid prototyping polymers for microfluidic devices and high
  pressure injections}.  \jt{Lab on a Chip}  \bvol{11},  \pg{3752--3765}.

\bibitem[Taylor(1953)]{Taylor1953Dispersion}
{\sc \au{Taylor, G.~I.}} \yr{1953}  \at{Dispersion of soluble matter in solvent
  flowing slowly through a tube}.  \jt{Proceedings of the Royal Society of
  London A}  \bvol{219},  \pg{186--203}.

\bibitem[Wang \& Christov(2019)]{Wang2019Theory}
{\sc \au{Wang, X.} \& \au{Christov, I.~C.}} \yr{2019}  \at{Theory of the
  flow-induced deformation of shallow compliant microchannels with thick
  walls}.  \jt{Proceedings of the Royal Society A}  \bvol{475},  \pg{20190513}.

\bibitem[Wang {\em et~al.\/}(2026)Wang, Zhang \& Qi]{Wang2026Control}
{\sc \au{Wang, X.}, \au{Zhang, M.} \& \au{Qi, H.}} \yr{2026}  \at{Control of
  axial dispersion in polymeric flows: Magnetic effect versus electro-osmotic
  effect}.  \jt{Journal of Fluid Mechanics}  \bvol{1045},  \pg{A19}.

\end{thebibliography}


\begin{thebibliography}{10}
\expandafter\ifx\csname natexlab\endcsname\relax\def\natexlab#1{#1}\fi
\def\au#1{#1} \def\ed#1{#1} \def\yr#1{#1}\def\at#1{#1}\def\jt#1{\textit{#1}}
  \def\bt#1{#1}\def\bvol#1{\textbf{#1}} \def\vol#1{#1} \def\pg#1{#1}
  \def\publ#1{#1}\def\arxiv#1{#1}\def\org#1{#1}\def\st#1{\textit{#1}}

\bibitem[Afonso {\em et~al.\/}(2009)Afonso, Alves \&
  Pinho]{Afonso2009Analytical}
{\sc \au{Afonso, A.~M.}, \au{Alves, M.~A.} \& \au{Pinho, F.~T.}} \yr{2009}
  \at{Analytical solution of mixed electro-osmotic/pressure driven flows of
  viscoelastic fluids in microchannels}.  \jt{Journal of Non-Newtonian Fluid
  Mechanics}  \bvol{159}~(1-3),  \pg{50--63}.

\bibitem[Aris(1956)]{Aris1956}
{\sc \au{Aris, R.}} \yr{1956}  \at{On the dispersion of a solute in a fluid
  flowing through a tube}.  \jt{Proceedings of the Royal Society of London A}
  \bvol{235},  \pg{67--77}.

\bibitem[Chandler \& Vella(2020)]{Chandler2020Validity}
{\sc \au{Chandler, T. G.~J.} \& \au{Vella, D.}} \yr{2020}  \at{Validity of
  {Winkler}'s mattress model for thin elastomeric layers: beyond {Poisson}'s
  ratio}.  \jt{Proceedings of the Royal Society A}  \bvol{476}~(2242),
  \pg{20200551}.

\bibitem[Chatwin(1970)]{Chatwin1970Normality}
{\sc \au{Chatwin, P.~C.}} \yr{1970}  \at{The approach to normality of the
  concentration distribution of a solute in a solvent flowing along a straight
  pipe}.  \jt{Journal of Fluid Mechanics}  \bvol{43}~(2),  \pg{321--352}.

\bibitem[Datta \& Ghosal(2008)]{Datta2008Dispersion}
{\sc \au{Datta, S.} \& \au{Ghosal, S.}} \yr{2008}  \at{Dispersion due to wall
  interactions in microfluidic separation systems}.  \jt{Physics of Fluids}
  \bvol{20}~(1),  \pg{012103}.

\bibitem[Dutta \& Beskok(2001)]{Dutta2001Combined}
{\sc \au{Dutta, P.} \& \au{Beskok, A.}} \yr{2001}  \at{Analytical solution of
  combined electroosmotic/pressure driven flows in two-dimensional straight
  channels: finite {Debye} layer effects}.  \jt{Analytical Chemistry}
  \bvol{73}~(9),  \pg{1979--1986}.

\bibitem[Gardiner(2009)]{Gardiner2009Stochastic}
{\sc \au{Gardiner, C.}} \yr{2009} {\em Stochastic Methods: A Handbook for the
  Natural and Social Sciences\/}, 4th edn.  \publ{Berlin, Heidelberg:
  Springer}.

\bibitem[Ghosal(2006)]{Ghosal2006Electrokinetic}
{\sc \au{Ghosal, S.}} \yr{2006}  \at{Electrokinetic flow and dispersion in
  capillary electrophoresis}.  \jt{Annual Review of Fluid Mechanics}
  \bvol{38},  \pg{309--338}.

\bibitem[Skotheim \& Mahadevan(2005)]{Skotheim2005Soft}
{\sc \au{Skotheim, J.~M.} \& \au{Mahadevan, L.}} \yr{2005}  \at{Soft
  lubrication: The elastohydrodynamics of nonconforming and conforming
  contacts}.  \jt{Physics of Fluids}  \bvol{17}~(9),  \pg{092101}.

\bibitem[Taylor(1953)]{Taylor1953Dispersion}
{\sc \au{Taylor, G.~I.}} \yr{1953}  \at{Dispersion of soluble matter in solvent
  flowing slowly through a tube}.  \jt{Proceedings of the Royal Society of
  London A}  \bvol{219},  \pg{186--203}.

\end{thebibliography}

\end{document}


\maketitle

\noindent

This supplementary material provides the derivations that underpin the reduced-order model presented in the manuscript. These include the local elastic-foundation wall law, the integration and asymptotic limits of the rigid-channel electroosmotic solution, the closed-form mixed-flow solution, the pressure-free Taylor–Aris dispersion coefficient, the Newtonian and rigid-channel consistency limits, the moment description of the band shape, the thin-EDL matched asymptotics, the multiple-scales reduction for a slowly varying gap, and the advection-dominated transit-time optimisation. The dimensional reference properties and parameter ranges are summarised in \S\ref{S-app:params}. Equations, figures, tables and sections prefixed by \textsf{S} refer to this supplementary material; all others refer to the main manuscript.

\vspace{6pt}

\section{Local elastic-foundation law}\label{S-app:wall}

Each compliant wall is represented by a compressible elastic layer of thickness \(h_c\) bonded to a rigid support. Under local constrained compression, the axial variation occurs on the long scale \(L\), while the dominant strain is normal to the wall. The constrained modulus is \(M=2G+\lambda_s\); the thin compressible-layer approximation \citep{Skotheim2005Soft,Chandler2020Validity} gives
\begin{equation}
	p^*(x^*)=M\frac{h_1^*(x^*)}{h_c},
	\qquad
	h_1^*(x^*)=\frac{h_c}{2G+\lambda_s} p^*(x^*).
	\label{S-eq:app_wall_dimensional}
\end{equation}
Using \(p_0=\eta_0U_{\mathrm{hs}}L/h_0^2\), \(p=p^*/p_0\) and \(h_1=h_1^*/h_0\) gives
\begin{equation}
	h_1=C_mp,
	\qquad
	h=1+C_mp,
	\qquad
	C_m=\frac{h_c}{h_0}
	\frac{\eta_0U_{\mathrm{hs}}L}
	{h_0^2(2G+\lambda_s)}.
	\label{S-eq:app_wall_dimensionless}
\end{equation}
The wall law is linear, although the full dependence of the flow equations on \(h=1+C_mp\) is retained. Its validity requires \(\beta\delta=h_c/L\ll1\), \(\chi=C_m\max_x|p|\ll1\) and the constrained strain \(\chi/\beta\ll1\). For pressure variations on the scale \(L\), constrained compression dominates when \((h_c/L)^2\ll1-2\nu_s\), where \(\nu_s\) is the Poisson ratio of the coating.

\section{Rigid-channel electroosmotic solution}\label{S-app:rigid}

The undeformed pressure-free solution, with \(h=1\), \(p'=0\) and \(E_x=1\), is given in Appendix~\ref{app:reference_state} of the main text. Here we record the integration steps and asymptotic limits within the Debye--H\"uckel and lubrication approximations used for sPTT electroosmosis \citep{Afonso2009Analytical}.

\subsection{Reference-flow integration}\label{S-app:reference_integration}

The axial momentum balance gives
\begin{equation}
	\frac{\mathrm{d}\tau_{xy}^{(0,0)}}{\mathrm{d}y}
	=-K^2\Psi(y),
	\qquad
	\tau_{xy}^{(0,0)}(0)=0.
\end{equation}
Integration yields
\begin{equation}
	\tau_{xy}^{(0,0)}(y)
	=-K\frac{\sinh(Ky)}{\cosh K}.
\end{equation}
The dimensionless sPTT shear relation is
\begin{equation}
	\frac{\mathrm{d}u^{(0,0)}}{\mathrm{d}y}
	=\tau_{xy}^{(0,0)}
	+\frac{2g}{K^2}
	\left(\tau_{xy}^{(0,0)}\right)^3.
\end{equation}
Applying $u^{(0,0)}(\pm1)=0$ gives the velocity profile \eqref{eq:u00}. Integration across the gap gives the flow rate \eqref{eq:Q00}, with the rheological correction \eqref{eq:Q1}.

\subsection{Asymptotic limits}\label{S-app:asymptotics}

Expansion of \eqref{eq:Q1} gives
\begin{align}
	\mathcal Q_1(K)
	&=\frac{4}{5}K^4-\frac{32}{35}K^6
	+O(K^8),
	&&K\to0,
	\label{S-eq:app_Q1_small}\\
	\mathcal Q_1(K)
	&=\frac{4}{3}-\frac{4}{9K}
	+O\!\left(e^{-2K}\right),
	&&K\to\infty.
	\label{S-eq:app_Q1_large}
\end{align}
Consequently,
\begin{align}
	Q^{(0,0)}
	&=\frac{2}{3}K^2
	+\left(\frac{4g}{5}-\frac{4}{15}\right)K^4
	+O(K^6),
	&&K\to0,\\
	Q^{(0,0)}
	&=2+\frac{4g}{3}
	-\frac{1}{K}\left(2+\frac{4g}{9}\right)
	+O\!\left(e^{-2K}\right),
	&&K\to\infty
\end{align}
along the fixed-\(g\) path.

\section{Closed-form mixed-flow relation}\label{S-app:mixed}

The nonlinear coupling of electroosmotic and pressure-driven shear follows the sPTT formulation of \citet{Afonso2009Analytical}. Let
\begin{equation}
	\alpha_p=p'(x),
	\qquad
	\alpha_E=\frac{E_xK}{\cosh(Kh)},
	\qquad
	a=\frac{2g}{K^2},
\end{equation}
so that
\begin{equation}
	\tau_{xy}=\alpha_py-\alpha_E\sinh(Ky).
\end{equation}
Expanding \(\tau_{xy}^3\) in \eqref{eq:Qexact} of the main text gives
\begin{multline}
	Q
	=2E_x\left[h-\frac{\tanh(Kh)}{K}\right]
	-\frac{2h^3}{3}\alpha_p\\
	-2a\left[
	\alpha_p^3I_1
	-3\alpha_p^2\alpha_EI_2
	+3\alpha_p\alpha_E^2I_3
	-\alpha_E^3I_4
	\right],
	\label{S-eq:app_mixed_flux}
\end{multline}
where
\begin{align}
	I_1&=\frac{h^5}{5},\\
	I_2&=
	\left(\frac{h^3}{K}+\frac{6h}{K^3}\right)\cosh(Kh)
	-\left(\frac{3h^2}{K^2}+\frac{6}{K^4}\right)\sinh(Kh),\\
	I_3&=
	\frac{h^2}{4K}\sinh(2Kh)
	-\frac{h}{4K^2}\cosh(2Kh)
	+\frac{1}{8K^3}\sinh(2Kh)
	-\frac{h^3}{6},\\
	I_4&=
	\frac{1}{4}
	\left[
	\frac{h}{3K}\cosh(3Kh)
	-\frac{1}{9K^2}\sinh(3Kh)
	-\frac{3h}{K}\cosh(Kh)
	+\frac{3}{K^2}\sinh(Kh)
	\right].
\end{align}
The \(p'^3\) and \(E_x^3\) terms are the pure pressure-driven and electroosmotic viscoelastic corrections, whereas \(p'^2E_x\) and \(p'E_x^2\) are the nonlinear mixed contributions. Setting \(g=0\) recovers the Newtonian mixed-flow relation; setting \(p'=0\), \(h=1\) and \(E_x=1\) recovers \eqref{eq:Q00}.

\section{Pressure-free Taylor--Aris coefficient}\label{S-app:TA}

Applying the Taylor--Aris cell problem \citep{Taylor1953Dispersion,Aris1956} to the rigid pressure-free profile, write
\begin{equation}
	u^{(0,0)}=u_N+g u_1,
	\qquad
	\varphi=\varphi_N+g \varphi_1,
\end{equation}
where
\begin{equation}
	-\varphi_j''
	=u_j-\langle u_j\rangle_y,
	\qquad
	\varphi_j'(\pm1)=0,
	\qquad
	\langle\varphi_j\rangle_y=0,
	\qquad j=N,1.
\end{equation}
Here \(\langle f\rangle_y=\tfrac12\int_{-1}^{1}f \mathrm{d}y\). The exact coefficient is quadratic in \(g\):
\begin{equation}
	\mathcal K_0
	=\mathcal K_{0,N}
	+g \mathcal K_{0,1}
	+g^2\mathcal K_{0,2},
	\label{S-eq:app_K_quadratic}
\end{equation}
with
\begin{equation}
	\mathcal K_{0,N}
	=\langle(\varphi_N')^2\rangle_y,
	\qquad
	\mathcal K_{0,1}
	=2\langle\varphi_N'\varphi_1'\rangle_y,
	\qquad
	\mathcal K_{0,2}
	=\langle(\varphi_1')^2\rangle_y.
\end{equation}
Direct integration gives
\begin{equation}
	\mathcal K_{0,N}
	=
	\frac{
	2K^2\cosh(2K)-8K^2-9K\sinh(2K)
	+12\cosh(2K)-12
	}{
	12K^4\cosh^2K
	},
	\label{S-eq:app_KN}
\end{equation}
and
\begin{multline}
	\mathcal K_{0,1}
	=
	\frac{1}{648K^4\cosh^4K}
	\bigl[
	24K^2\cosh^2(2K)
	-336K^2\cosh(2K)
	+1284K^2\\
	+1416K\sinh(2K)
	-39K\sinh(4K)
	+80\cosh^2(2K)
	-1984\cosh(2K)
	+1904
	\bigr].
	\label{S-eq:app_K1}
\end{multline}
The positive quadratic term \(\mathcal K_{0,2}\) is evaluated directly from its defining integral. Its leading limits, together with those of \(\mathcal K_{0,N}\) and \(\mathcal K_{0,1}\), are
\begin{align}
	\mathcal K_{0,N}
	&\sim\frac{2K^4}{945},
	&
	\mathcal K_{0,1}
	&\sim\frac{16K^6}{4725},
	&
	\mathcal K_{0,2}
	&\sim\frac{8K^8}{5775},
	&&K\to0,
	\\
	\mathcal K_{0,N}
	&\sim\frac{1}{3K^2},
	&
	\mathcal K_{0,1}
	&\sim\frac{4}{27K^2},
	&
	\mathcal K_{0,2}
	&\sim\frac{4}{243K^2},
	&&K\to\infty.
\end{align}
Thus, at fixed \(g\),
\begin{equation}
	\mathcal K_0
	\sim
	\frac{2K^4}{945},
	\qquad K\to0,
\end{equation}
whereas
\begin{equation}
	\mathcal K_0
	\sim
	\frac{1}{K^2}
	\left(
	\frac13+\frac{4g}{27}+\frac{4g^2}{243}
	\right),
	\qquad K\to\infty.
\end{equation}
For the parabolic profile \(u=1-y^2\), the same cell formulation gives the classical result \(\mathcal K=8/945\).

\section{Newtonian and rigid-channel limits}\label{S-app:limits}

In the Newtonian limit \(g=0\) \citep{Dutta2001Combined},
\begin{equation}
	u^{(0,0)}
	=1-\frac{\cosh(Ky)}{\cosh K},
	\qquad
	Q^{(0,0)}
	=2\left(1-\frac{\tanh K}{K}\right).
\end{equation}
For a rigid Newtonian slit at stall,
\begin{equation}
	p_{\mathrm{stall}}
	=3\left(1-\frac{\tanh K}{K}\right)
	=\frac32Q^{(0,0)}.
\end{equation}
Hence \(p_{\mathrm{stall}}\sim K^2\) as \(K\to0\) and \(p_{\mathrm{stall}}\to3\) as \(K\to\infty\). For a uniform pressure-free channel, \(h=1\) and both operating modes give \(E_x=1\). In a deformed channel,
\begin{equation}
	E_x^{\mathrm{CC}}=h^{-1},
	\qquad
	E_x^{\mathrm{CV}}
	=\frac{h^{-1}}{\langle h^{-1}\rangle_x},
\end{equation}
so the operating modes differ through the normalisation of the conserved product \(hE_x\), while pressure generation is selected by the hydraulic boundary condition.

\section{Moment description of the band shape}\label{S-app:moments}

For an axially unbounded rigid slit of uniform half-height \(h_0\), introduce the axial coordinate \(\xi\) in a frame moving with the mean velocity and scale length and time with \(h_0\) and \(h_0^2/D_s^*\), respectively. The transport equation is
\begin{equation}
	\partial_tc
	+\mathrm{Pe} w(y) \partial_\xi c
	=\partial_{\xi\xi}c+\partial_{yy}c,
	\qquad
	w(y)=u(y)-\bar u.
\end{equation}
Following the method of moments of \citet{Aris1956}, define the axial moments
\begin{equation}
	c_n(y,t)
	=\int_{-\infty}^{\infty}\xi^n c(\xi,y,t) \mathrm{d}\xi.
\end{equation}
Multiplication by \(\xi^n\) and integration over \(\xi\) give
\begin{equation}
	\partial_tc_n
	=\partial_{yy}c_n
	+n \mathrm{Pe} w c_{n-1}
	+n(n-1)c_{n-2},
	\qquad
	\partial_yc_n(\pm1,t)=0.
	\label{S-eq:app_moment_hierarchy}
\end{equation}
For a normalised pulse in the comoving frame, the long-time transverse corrections are described by
\begin{align}
	B_1''&=-w,
	&
	B_1'(\pm1)&=0,
	&
	\langle B_1\rangle_y&=0,\\
	B_2''&=-\left(wB_1-\langle wB_1\rangle_y\right),
	&
	B_2'(\pm1)&=0,
	&
	\langle B_2\rangle_y&=0.
\end{align}
The variance and third central moment then grow as
\begin{equation}
	\frac{\mathrm{d}\mu_2}{\mathrm{d}t}
	=2\left(1+\mathrm{Pe}^2\mathcal K\right),
	\qquad
	\frac{\mathrm{d}\mu_3}{\mathrm{d}t}
	=6\mathrm{Pe}^3\mathcal D_3,
\end{equation}
where
\begin{equation}
	\mathcal K
	=\langle wB_1\rangle_y
	=\langle(B_1')^2\rangle_y,
	\qquad
	\mathcal D_3=\langle wB_2\rangle_y.
\end{equation}
After the mean band has travelled the reference distance \(L\), the leading spatial skewness is
\begin{equation}
	\gamma_1\sim C_SN_{\mathrm{sp}}^{-1/2},
	\qquad
	C_S
	=
	\frac{
	3\mathrm{Pe}^4\bar u \mathcal D_3
	}{
	2\left(1+\mathrm{Pe}^2\mathcal K\right)^2
	}.
	\label{S-eq:app_skewness}
\end{equation}
Here \(N_{\mathrm{sp}}=\bar u/(2D_{\mathrm{ax}})\) is the spatial sharpness measure, with \(D_{\mathrm{ax}}\) expressed in the main-text scaling of \eqref{eq:effective_drift}. The excess kurtosis enters at the next order and is \(O(N_{\mathrm{sp}}^{-1})\), consistent with the long-time approach to normality \citep{Chatwin1970Normality}. For the loaded profiles considered in figure~\ref{fig:skew} of the main text, \(\mathcal D_3<0\), corresponding to a rear-tailed band.

\section{Thin-EDL matched asymptotics}\label{S-app:slip}

\subsection{EDL matching and effective slip}

The matching follows the thin-EDL slip construction described by \citet{Ghosal2006Electrokinetic}, with the sPTT shear law \citep{Afonso2009Analytical} retained within the charged layer. We take \(K\to\infty\) at fixed \(g\), with bounded \(h\), \(E_x\) and \(p'\), and a positive gap. The fields below denote their leading limiting values unless a finite-\(K\) expansion is stated. Consider the upper wall and introduce the inner coordinate \(Y=K[h(x)-y]\). For \(Kh\gg1\), the local Debye--H\"uckel potential is
\begin{equation}
	\Psi_{\mathrm{in}}=e^{-Y}.
\end{equation}
The electroosmotic contribution to the shear stress is
\begin{equation}
	\tau_{xy}^{\mathrm{in}}
	\sim-KE_xe^{-Y},
\end{equation}
while the pressure contribution is \(O(1)\) and is subdominant within the EDL. At fixed \(g\), the sPTT shear relation gives
\begin{equation}
	\frac{\mathrm{d}u_{\mathrm{in}}}{\mathrm{d}Y}
	=E_xe^{-Y}
	+2gE_x^3e^{-3Y}.
\end{equation}
No slip at \(Y=0\) therefore yields
\begin{equation}
	u_{\mathrm{in}}(Y)
	=E_x\left(1-e^{-Y}\right)
	+\frac{2g}{3}E_x^3\left(1-e^{-3Y}\right).
	\label{S-eq:app_inner_velocity}
\end{equation}
Matching to the outer core gives the effective slip
\begin{equation}
	U_s(x)
	=E_x(x)+\frac{2g}{3}E_x^3(x).
	\label{S-eq:app_effective_slip}
\end{equation}
The displacement deficit of either EDL is
\begin{equation}
	\frac{1}{K}
	\int_0^\infty
	\left(U_s-u_{\mathrm{in}}\right) \mathrm{d}Y
	=
	\frac{1}{K}
	\left(
	E_x+\frac{2g}{9}E_x^3
	\right).
\end{equation}
Thus, the local pressure-free electroosmotic flow rate satisfies
\begin{equation}
	Q_{\mathrm{EOF}}
	=
	2hU_s
	-\frac{2}{K}
	\left(
	E_x+\frac{2g}{9}E_x^3
	\right)
	+o(K^{-1}).
	\label{S-eq:app_thin_EDL_flux}
\end{equation}
For \(h=E_x=1\), \eqref{S-eq:app_thin_EDL_flux} reproduces the large-\(K\) expansion of \eqref{eq:Q00}. In the loaded outer problem, \(U_s(x)\) supplies the effective wall condition for the pressure-driven core, leading to \eqref{eq:thin_EDL_velocity} and \eqref{eq:anchor} of the main text.

\subsection{Core cell problem and the loaded plateau}
\label{S-sec:core_plateau}

Since \(a=2g/K^2\to0\), the leading outer shear relation is Newtonian even when \(g>0\). Define the pressure-driven mean-velocity deficit
\begin{equation}
\Delta_u(x)=U_s(x)-\bar u(x)=\frac{p'(x)h^2(x)}{3}.
\end{equation}
The core velocity deviation and cell gradient are
\begin{equation}
u-\bar u=\frac{\Delta_u}{2}\left(\frac{3y^2}{h^2}-1\right),
\qquad
\partial_y\varphi=\frac{\Delta_u}{2}\left(y-\frac{y^3}{h^2}\right).
\label{S-eq:core_cell_gradient}
\end{equation}
The second expression follows by integrating \(-\partial_{yy}\varphi=u-\bar u\) from the symmetry plane and satisfies the wall Neumann conditions. Thus,
\begin{equation}
\mathcal K_\infty
=\frac{\Delta_u^2}{4h}\int_0^h
\left(y-\frac{y^3}{h^2}\right)^2 \mathrm{d}y
=\frac{2}{105}h^2\Delta_u^2.
\label{S-eq:core_plateau}
\end{equation}
For a rigid uniform channel, \(Q_{\mathrm{free}}\to2hU_s\) and \(\bar u\to rU_s\); hence \(\Delta_u=U_s(1-r)\) and the plateau scales as \((1-r)^2\). This scaling reflects the quadratic cell energy, rather than a requirement for local flow reversal. In the rigid reference slit, \(h=E_x=1\), put \(S=1+2g/3\). The limiting centreline velocity is \(S(3r-1)/2\), so it vanishes at \(r=1/3\) and is negative for \(r<1/3\). Core shear persists for all \(r<1\).

\subsection{Compliance under constant current and constant voltage}
\label{S-sec:compliant_plateau}

Let \(E_x=\mathcal J/h\), with \(\mathcal J=1\) under CC operation and \(\mathcal J\int_0^1h^{-1} \mathrm{d}x=1\) under CV operation. The undeformed free-flow branch has \(Q_{\mathrm{free}}\to2S\) in both modes. At fixed throughput fraction, define
\begin{equation}
F(x)=h(U_s-\bar u)
=\mathcal J+\frac{2g\mathcal J^3}{3h^2(x)}-rS.
\end{equation}
Then the limiting pressure, geometry and dispersion satisfy
\begin{equation}
p'=\frac{3F}{h^3},\qquad
h'=\frac{3C_mF}{h^3},\qquad
h(0)=1,\qquad
\mathcal K_\infty=\frac{2F^2}{105}.
\label{S-eq:limiting_pump_system}
\end{equation}
This is a closed limiting pump problem, supplemented by the voltage constraint for CV operation. On the positive-loading branch \(F>0\). For a Newtonian fluid under CC operation, \(F=1-r\), so the plateau is independent of the deformed gap; indeed, \(h^4=1+12C_m(1-r)x\). For \(g>0\), the CC value of \(F\) decreases as \(h\) increases, lowering the local plateau relative to the rigid reference at fixed \(r\) and \(g\). The normalisation \(\mathcal J>1\) under CV operation changes this balance. To isolate the first compliance correction, set \(q=1-r>0\) and expand at fixed \(g\) and \(r\) as \(C_m\to0\). Both operating modes have the same leading pressure and gap:
\begin{equation}
p_0(x)=3Sqx,\qquad
h(x)=1+3C_mSqx+O(C_m^2).
\end{equation}
The conserved field constant is
\begin{equation}
\mathcal J_{\mathrm{CC}}=1,\qquad
\mathcal J_{\mathrm{CV}}
=1+\frac{3}{2}C_mSq+O(C_m^2).
\end{equation}
Writing \(\mathcal K_{\mathrm r}=2S^2q^2/105\) for the rigid plateau gives
\begin{align}
\frac{\mathcal K_\infty^{\mathrm{CC}}(x)}{\mathcal K_{\mathrm r}}
&=1-8gC_mx+O(C_m^2),\\
\frac{\mathcal K_\infty^{\mathrm{CV}}(x)}{\mathcal K_{\mathrm r}}
&=1+C_m(3+6g-8gx)+O(C_m^2).
\label{S-eq:local_compliance_correction}
\end{align}
Their path averages therefore obey
\begin{equation}
\frac{\overline{\mathcal K}_\infty^{\mathrm{CC}}}{\mathcal K_{\mathrm r}}
=1-4gC_m+O(C_m^2),\qquad
\frac{\overline{\mathcal K}_\infty^{\mathrm{CV}}}{\mathcal K_{\mathrm r}}
=1+(3+2g)C_m+O(C_m^2).
\label{S-eq:mean_compliance_correction}
\end{equation}
These expansions concern the thin-EDL shear coefficient at prescribed \(r<1\), not a fixed back pressure or resistance. They require weak compliance at fixed \(g\), together with small deformation and wall strain. At \(r=1\), the state is undeformed and both leading plateaux vanish; the above normalised ratios are then undefined. Changes in the first-passage plate number also involve the mean speed and effective drift and cannot be inferred from \eqref{S-eq:mean_compliance_correction} alone.

\subsection{Weak-loading crossover and shear cancellation}
\label{S-sec:weak_loading}

Consider the rigid reference slit \(h=E_x=1\) and define
\begin{equation}
S=1+\frac{2g}{3},\qquad
B=1+\frac{2g}{9},\qquad
q=1-r.
\end{equation}
The inner solution gives an integrated velocity deficit \(B/K\) at either wall, and \(Q_{\mathrm{free}}/2=S-B/K+O(e^{-2K})\). Thus the pressure-free cell coefficient is \(B^2/(3K^2)+O(K^{-3})\), consistent with \S S4. The loaded plateau is \(2S^2q^2/105\); equating these two positive contributions yields the balance scale
\begin{equation}
K_c=\sqrt{\frac{35}{2}} \frac{B}{Sq}.
\label{S-eq:crossover_balance}
\end{equation}
The scale alone does not describe the cancellation between the two cell responses. To retain that interaction, take the distinguished limit \(q=O(K^{-1})\) at fixed \(g\). The mixed-flow relation gives \(p'=3Sq+O(K^{-2})\). Away from the EDLs, the velocity deviation and its integral on the upper half-gap are
\begin{align}
u-\bar u
&=\frac{B}{K}+\frac{Sq}{2}(3y^2-1)+O(K^{-2}),\\
\int_0^y(u-\bar u) \mathrm{d}s
&=\frac{B}{K}y+\frac{Sq}{2}(y^3-y)+O(K^{-2}).
\label{S-eq:weak_load_cell_integral}
\end{align}
The EDL restores the zero cell gradient at the wall. Its width is \(O(K^{-1})\) and the cell gradient there is \(O(K^{-1})\), so its contribution to the error in the cell energy is \(O(K^{-3})\). Squaring the leading integral and averaging gives
\begin{align}
\mathcal K
&=\int_0^1\left[\frac{B}{K}y+
\frac{Sq}{2}(y^3-y)\right]^2 \mathrm{d}y+O(K^{-3})\nonumber\\
&=\frac{B^2}{3K^2}-\frac{2SBq}{15K}
+\frac{2S^2q^2}{105}+O(K^{-3}).
\label{S-eq:weak_load_dispersion}
\end{align}
The negative cross term follows from \(y(y^3-y)<0\) for \(0<y<1\). It expresses the opposing signs of the EDL-deficit and pressure-driven cell responses. With \(z=SKq/B\), the corresponding scaled form is
\begin{equation}
\frac{K^2\mathcal K}{B^2}
=\frac13-\frac{2z}{15}+\frac{2z^2}{105}+O(K^{-1}),
\qquad z=O(1).
\label{S-eq:weak_load_scaled}
\end{equation}
At prescribed small \(q\), minimising the leading expression in \eqref{S-eq:weak_load_dispersion} with respect to \(K\) gives
\begin{equation}
K_{\min}\sim\frac{5B}{Sq},\qquad
\mathcal K_{\min}\sim\frac{S^2q^2}{175}
=\frac{3}{10}\mathcal K_{\mathrm r},\qquad q\to0.
\label{S-eq:weak_load_minimum}
\end{equation}
The balance scale \(K_c\) and the minimum location are therefore distinct, although both diverge as \(q^{-1}\). These formulas are asymptotic near free flow and are not used to predict the reference-case minimum at \(r=0.5\). The plate-number optimum additionally depends on the mean velocity and molecular diffusion. For a compliant channel, neither the rigid crossover prefactor nor \eqref{S-eq:weak_load_scaled} is universal; the local slip and self-consistent gap must also be retained.

\subsection{Free-flow and stall endpoints}

As \(r\to1\), the rigid plateau vanishes as \(q^2\), while \(K_c\) diverges as \(q^{-1}\). The unscaled coefficient tends to zero in either order of the limits \(K\to\infty\) and \(r\to1\). The non-uniformity lies in the leading scaling, which depends on \(Kq\), rather than in different values of these limits. At \(r=0\), the rigid thin-EDL velocity is \(u=S(3y^2-1)/2\) and \(\mathcal K_\infty=2S^2/105\). Although the mean flow vanishes, particles still sample the opposing velocities by transverse diffusion. In the uniform reduced first-passage problem \citep{Gardiner2009Stochastic}, put \(D=D_{\mathrm{ax}}>0\). The reflecting-inlet and absorbing-outlet conditions give
\begin{equation}
T(x)=\frac{1-x^2}{2D},\qquad
V(x)=\frac{1-x^4}{6D^2},\qquad
N=\frac32.
\label{S-eq:uniform_diffusive_stall}
\end{equation}
This is a diffusion-controlled exit statistic, not finite-throughput separation. Compliance makes the area and effective coefficients non-uniform, so the stall value need not be \(3/2\). Molecular diffusion is retained throughout. Finally, fixed \(g=\epsilon_p(\lambda\kappa U_{\mathrm{hs}})^2\) is an essential specification of the asymptotic path. If \(\kappa\) alone increases at fixed \(\epsilon_p\), \(\lambda\), \(U_{\mathrm{hs}}\) and \(h_0\), then \(g/K^2\) remains fixed and the outer cubic constitutive term does not vanish. That limit requires a different analysis. Any dimensional interpretation must also satisfy the lubrication, constitutive, electrostatic and transport-validity conditions stated in the main text.

\section{Multiple-scales reduction for a slowly varying gap}\label{S-app:varea}

The transverse-equilibration argument underlying Taylor--Aris dispersion \citep{Taylor1953Dispersion,Aris1956} is applied here to the slowly varying gap, following the long-time reduction for axially varying cross-sections \citep{Datta2008Dispersion}. The dimensionless advection--diffusion equation is
\begin{equation}
	\partial_tc
	+u \partial_xc
	+v \partial_yc
	=
	\frac{1}{\mathrm{Pe} \delta} 
	\partial_{yy}c
	+\frac{\delta}{\mathrm{Pe}} 
	\partial_{xx}c.
	\label{S-eq:app_transport_dimless}
\end{equation}
Introduce \(\eta=y/h(x)\), together with
\begin{equation}
	a(x,\eta)=\eta\frac{h'}{h},
	\qquad
	W(x,\eta)=v-u\eta h'.
\end{equation}
Then
\begin{equation}
	\left.\partial_x\right|_y
	=
	\left.\partial_x\right|_\eta
	-a \partial_\eta,
	\qquad
	\partial_y=h^{-1}\partial_\eta,
\end{equation}
and \eqref{S-eq:app_transport_dimless} becomes
\begin{equation}
	\partial_tc
	+u \partial_xc
	+\frac{W}{h} \partial_\eta c
	=
	\frac{1}{\mathrm{Pe} \delta h^2}
	\partial_{\eta\eta}c
	+\frac{\delta}{\mathrm{Pe}}
	\left(\partial_x-a\partial_\eta\right)^2c.
	\label{S-eq:app_mapped_transport}
\end{equation}
Continuity gives
\begin{equation}
	\partial_\eta W=-\partial_x(hu),
	\qquad
	W(x,0)=0.
\end{equation}
The exact wall no-flux condition reduces at leading lubrication order to \(\partial_\eta c=0\) at \(\eta=\pm1\); the omitted conormal correction is \(O(\delta^2hh'\partial_xc)\). Let \(\mu=\mathrm{Pe} \delta\ll1\) and expand
\begin{equation}
	c=\bar c+\mu c_1+O(\mu^2).
\end{equation}
At leading order, \(\bar c\) is independent of \(\eta\). At the next order,
\begin{equation}
	c_1=-h^2\widehat{\varphi} \partial_x\bar c,
\end{equation}
where
\begin{equation}
	-\partial_{\eta\eta}\widehat{\varphi}
	=u-\bar u,
	\qquad
	\partial_\eta\widehat{\varphi}(\pm1)=0,
	\qquad
	\langle\widehat{\varphi}\rangle_\eta=0,
\end{equation}
and
\begin{equation}
	\bar u
	=\langle u\rangle_\eta
	=\frac{Q}{2h},
	\qquad
	\langle f\rangle_\eta
	=\frac12\int_{-1}^{1}f \mathrm{d}\eta.
\end{equation}
The physical-coordinate corrector in \eqref{eq:cellloaded} of the main text is
\begin{equation}
	\varphi(x,y)
	=h^2(x)\widehat{\varphi}(x,\eta),
\end{equation}
and hence
\begin{equation}
	\mathcal K(x)
	=
	h^2
	\left\langle
	\left(\partial_\eta\widehat{\varphi}\right)^2
	\right\rangle_\eta
	=
	\left\langle
	\left(\partial_y\varphi\right)^2
	\right\rangle_y.
	\label{S-eq:app_mapped_K}
\end{equation}

The next solvability condition gives
\begin{equation}
	\frac{\partial}{\partial t}
	\left(A\bar c\right)
	+\frac{\partial}{\partial x}
	\left(Q\bar c\right)
	=
	\delta\frac{\partial}{\partial x}
	\left[
	A\left(
	\mathrm{Pe}^{-1}
	+\mathrm{Pe} \mathcal K
	\right)
	\frac{\partial\bar c}{\partial x}
	\right],
	\qquad
	A=2h,
	\label{S-eq:app_area_transport}
\end{equation}
up to the formal remainder
\(O\!\left(\mu^2+\delta|h'|\right)\) in the retained axial flux. The terms involving \(W\) are retained in the solvability calculation; the practical accuracy of \eqref{S-eq:app_area_transport} is assessed independently in \S\ref{sec:transport_validation} of the main text.

\subsection{First-passage equations}

Let
\begin{equation}
	n(x,t)=A(x)\bar c(x,t),
	\qquad
	D_{\mathrm{ax}}
	=\delta\left(
	\mathrm{Pe}^{-1}
	+\mathrm{Pe} \mathcal K
	\right).
\end{equation}
Equation \eqref{S-eq:app_area_transport} can be written as
\begin{equation}
	\partial_tn
	=-\partial_x(bn)
	+\partial_{xx}\left(D_{\mathrm{ax}}n\right),
\end{equation}
where
\begin{equation}
	b
	=\bar u+D_{\mathrm{ax}}'
	+D_{\mathrm{ax}}\frac{A'}{A}.
\end{equation}
The backward first-passage formulation \citep{Gardiner2009Stochastic} gives the equations for the mean \(T(x)\) and centred second moment \(V(x)\):
\begin{align}
	bT'+D_{\mathrm{ax}}T''&=-1,\\
	bV'+D_{\mathrm{ax}}V''
	&=-2D_{\mathrm{ax}}(T')^2,
\end{align}
with
\begin{equation}
	T'(0)=V'(0)=0,
	\qquad
	T(1)=V(1)=0.
\end{equation}
The outlet statistics are
\begin{equation}
	t_R=T(0),
	\qquad
	\sigma_t^2=V(0),
	\qquad
	N=\frac{t_R^2}{\sigma_t^2}.
\end{equation}
The full backward equations are used near stall.

\subsection{Advection-dominated moments and P\'eclet-number optimum}
\label{S-app:transit_optimisation}

For forward flow, $\bar u>0$, in the advection-dominated regime, $\max_x(D_{\mathrm{ax}}/\bar u)\ll1$, the leading balance in \eqref{eq:backward} gives
\begin{equation}
	T'\simeq-\bar u^{-1},
	\qquad
	V'\simeq-2D_{\mathrm{ax}}\bar u^{-3}.
\end{equation}
Integration from the absorbing outlet, with $T(1)=V(1)=0$, gives the transit-time integrals \eqref{eq:moments} in Appendix~\ref{app:transit_moments}. The leading outer solution applies away from the reflecting-inlet boundary layer and is not used near stall, where $\bar u\to0$. At fixed hydrodynamics, the leading mean transit time is independent of $\mathrm{Pe}$. Maximising the corresponding plate number is therefore equivalent to minimising its temporal variance. Separating the molecular and shear contributions in \eqref{eq:moments} gives
\begin{equation}
	\sigma_t^2\simeq2\delta
	\left(
	\frac{A_0}{\mathrm{Pe}}
	+\mathrm{Pe} A_1
	\right),
	\qquad
	A_0=\int_0^1\bar u^{-3} \mathrm{d}x,
	\qquad
	A_1=\int_0^1\mathcal K\bar u^{-3} \mathrm{d}x.
\end{equation}
For $A_1>0$, the stationary condition is $-A_0/\mathrm{Pe}^2+A_1=0$, giving \eqref{eq:Pe_optimum} with $\mathcal K_t=A_1/A_0$. The second derivative is $4\delta A_0/\mathrm{Pe}^3>0$, and the two contributions to the leading variance are equal at the optimum. The estimate assumes fixed hydrodynamics and advection-dominated transit. The full backward equations give the plotted reduced-model plate numbers.

\section{Dimensional reference properties and ranges}\label{S-app:params}

Table~\ref{S-tab:params} lists the dimensional reference properties and ranges used to map the dimensionless calculations in the main text. The dimensional ranges are linked by the physical relations in \S\ref{sec:regime_checks} of the main text. The quoted conductivity, concentration and Debye length are reference values at \(K=10\). The calculations include the sensitivity cases and bulk-Ohmic extrapolations identified in that section.

\begin{table}[htbp]
	\centering
	\caption{Dimensional reference properties and ranges for mapping the dimensionless calculations.}
	\label{S-tab:params}
	\begin{tabular}{llll}
		\toprule
		Symbol & Quantity & Reference value & Range considered \\
		\midrule
		$h_0$ & undeformed half-height
		& $1~\mu\mathrm{m}$ & $0.5$--$10~\mu\mathrm{m}$ \\
		$L$ & channel length
		& $1~\mathrm{mm}$ & $0.5$--$10~\mathrm{mm}$ \\
		$h_c$ & compliant-layer thickness
		& $2~\mu\mathrm{m}$ & $\beta=h_c/h_0=2$ \\
		$T_{\mathrm{abs}}$ & absolute temperature
		& $293~\mathrm{K}$ & $293$--$300~\mathrm{K}$ \\
		$\zeta_0$ & wall $\zeta$-potential
		& $-25~\mathrm{mV}$ & $-25~\mathrm{mV}\leq\zeta_0<0$ \\
		$E_0$ & reference electric field
		& $100~\mathrm{kV m^{-1}}$ & $10$--$100~\mathrm{kV m^{-1}}$ \\
		$\eta_0$ & zero-shear viscosity
		& $10^{-3}~\mathrm{Pa s}$ & fixed \\
		$\lambda$ & relaxation time
		& $29~\mu\mathrm{s}$ & $3$--$300~\mu\mathrm{s}$ \\
		$2G+\lambda_s$ & constrained modulus
		& $34.8~\mathrm{kPa}$ & $10$--$200~\mathrm{kPa}$ \\
		$\rho_f$ & fluid density
		& $10^3~\mathrm{kg m^{-3}}$ & $998$--$1000~\mathrm{kg m^{-3}}$ \\
		$\varepsilon_e$ & electrolyte permittivity
		& $695.4\times10^{-12}~\mathrm{F m^{-1}}$ & fixed \\
		$\sigma$ & bulk conductivity
		& $1.39\times10^{-4}~\mathrm{S m^{-1}}$ & linked to $\kappa$ \\
		$c_\infty$ & bulk ionic concentration
		& $9.1\times10^{-3}~\mathrm{mol m^{-3}}$ & linked to $\kappa$ and $T_{\mathrm{abs}}$ \\
		$D_s^*$ & solute diffusivity
		& $10^{-10}~\mathrm{m^2 s^{-1}}$ & $10^{-11}$--$10^{-9}~\mathrm{m^2 s^{-1}}$ \\
		$D_i^*$ & ionic diffusivity
		& $2\times10^{-9}~\mathrm{m^2 s^{-1}}$ & fixed \\
		$k_f$ & thermal conductivity
		& $0.6~\mathrm{W m^{-1} K^{-1}}$ & fixed \\
		$\kappa^{-1}$ & Debye length
		& $100~\mathrm{nm}$ & $50$--$500~\mathrm{nm}$ \\
		\bottomrule
	\end{tabular}
\end{table}

\bibliographystyle{jfm}
\bibliography{jfm}